\documentclass[aps,prb,superscriptaddress,reprint]{revtex4-2}

\usepackage[utf8]{inputenc}
\usepackage{graphicx,dcolumn,bm,amsmath,amssymb,euscript}

\graphicspath{{figures/BCF_ABsteps_GaN/}}

\newcommand{\kapt}{\tilde{\kappa}}
\newcommand{\kapsq}{\langle \kappa^2 \rangle}
\newcommand{\rhot}{\tilde{\rho}}
\newcommand{\pt}{\tilde{p}}
\newcommand{\qt}{\tilde{q}}

	\usepackage{color, colortbl}   
	\usepackage[bookmarksopen,bookmarksnumbered,colorlinks,citecolor=red,urlcolor=blue,filecolor=green]{hyperref}

\begin{document}

\title{Burton-Cabrera-Frank theory for surfaces with alternating step types}

\author{Guangxu Ju}
    \email[correspondence to: ]{juguangxu@gmail.com}
	\altaffiliation[current address: ]{Lumileds Lighting Co., San Jose, CA 95131 USA.}
	\affiliation{Materials Science Division, Argonne National Laboratory, Lemont, IL 60439 USA}
\author{Dongwei Xu}
	\affiliation{Materials Science Division, Argonne National Laboratory, Lemont, IL 60439 USA}
	\affiliation{School of Energy and Power Engineering, Huazhong University of Science and Technology, Wuhan, Hubei 430074, China}
\author{Carol Thompson}
	\affiliation{Department of Physics, Northern Illinois University, DeKalb, IL 60115 USA}
\author{Matthew J. Highland}
	\affiliation{X-ray Science Division, Argonne National Laboratory, Lemont, IL 60439 USA}
\author{Jeffrey A. Eastman}
	\affiliation{Materials Science Division, Argonne National Laboratory, Lemont, IL 60439 USA}
\author{Weronika Walkosz}
	\affiliation{Department of Physics, Lake Forest College, Lake Forest, IL 60045 USA}
\author{Peter Zapol}
	\affiliation{Materials Science Division, Argonne National Laboratory, Lemont, IL 60439 USA}
\author{G. Brian Stephenson}
    \email[correspondence to: ]{stephenson@anl.gov}
	\affiliation{Materials Science Division, Argonne National Laboratory, Lemont, IL 60439 USA}

\date{January 26, 2022}  

\begin{abstract}
Burton-Cabrera-Frank (BCF) theory has proven to be a versatile framework to relate surface morphology and dynamics during crystal growth to the underlying mechanisms of adatom diffusion and attachment at steps.
For an important class of crystal surfaces, including the basal planes of hexagonal close-packed and related systems, the steps in a sequence on a vicinal surface can exhibit properties that alternate from step to step.
Here we develop BCF theory for such surfaces, relating observables such as alternating terrace widths as a function of growth conditions to the kinetic coefficients for adatom attachment at steps.
We include the effects of step transparency and step-step repulsion.
A general solution is obtained for the dynamics of the terrace widths, assuming quasi-steady-state adatom distributions on the terraces.
An explicit simplified analytical solution is obtained under widely applicable approximations.
From this we obtain expressions for the full-steady-state terrace fraction as a function of growth rate.
Fits of the theoretical predictions to recent experimental determinations of the steady-state and dynamics of terrace fractions on GaN (0001) surfaces during organo-metallic vapor phase epitaxy give values of the kinetic coefficients for this system.
In Appendices, we also connect a model for diffusion between kinks on steps to the model for diffusion between steps on terraces, which quantitatively relates step transparency to the kinetics of atom attachment at kinks,
and consider limiting cases of diffusion-limited, attachment-limited, and mixed kinetics.
\end{abstract}

\maketitle
\section{Introduction}

The atomic-scale mechanisms of crystal growth are often described within the framework of Burton-Cabrera-Frank (BCF) theory \cite{1951_Burton_PhilTransRS243_29,1988_Ghez_IBMJRD32_804,1999_Jeong_SurfSciRep34_171,2005_Krug_ISNM149_69,2015_Woodruff_PhilTransA373_20140230}, in which deposited adatoms diffuse on top of the exposed atomic layers (terraces) of the crystal surface, until they either attach to existing steps at terrace edges, join together to nucleate a new terrace, or evaporate.
By matching adatom diffusion on terraces to flux boundary conditions associated with adatom attachment kinetics at the steps defining the terrace edges, BCF theory provides a detailed phenomenological description that is often used to analyze step-flow growth.
The steps can interact not only through the adatom diffusion field but also through terrace-width-dependent adatom chemical potentials that characterize elastic and entropic effects \cite{1999_Jeong_SurfSciRep34_171,2010_Patrone_PRE82_061601}.
While BCF theory can be formulated to consider two-dimensional diffusion, e.g. to model the meandering of curved steps \cite{bales1990morphological,1994_Saito_PRB49_10677,2000_Gillet_EurPhysJB18_519,2003_Pierre-Louis_PRE68_021604,2007_Sato_EurPhysJB59_311}, the simple case of one-dimensional diffusion between relatively straight steps on vicinal surfaces is also very powerful.
One-dimensional BCF models have been used extensively to understand the step-bunching instability \cite{2020_Guin_PRL124_036101,2003_Pierre-Louis_SurfSci529_114,2017_Bellmann_JCrystGrowth478_187,2007_Dufay_PRB75_241304,2016_Li_ApplSurfSci371_242,2000_Pimpinelli_SurfSci445_L23,sato1995morphological}, step pairing \cite{2004_Pierre-Louis_PRL93_165901,sato1997kinematical}, step width fluctuations \cite{2010_Patrone_PRE82_061601}, growth mode transitions \cite{2007_Ranguelov_PRB75_245419}, and effects of surface chemistry \cite{2000_Pimpinelli_SurfSci445_L23,2008_Chua_APL92_013117,2019_Hanada_PhysRevMater3_103404,2020_Redkov_CGD20_2590}. 
The parameters in BCF models can be related to those in kinetic Monte Carlo models for surface dynamics \cite{2014_Patrone_MultModSim12_364, 2011_Zaluska-Kotur_JAP109_023515}.

\begin{figure}
\includegraphics[width=\linewidth]{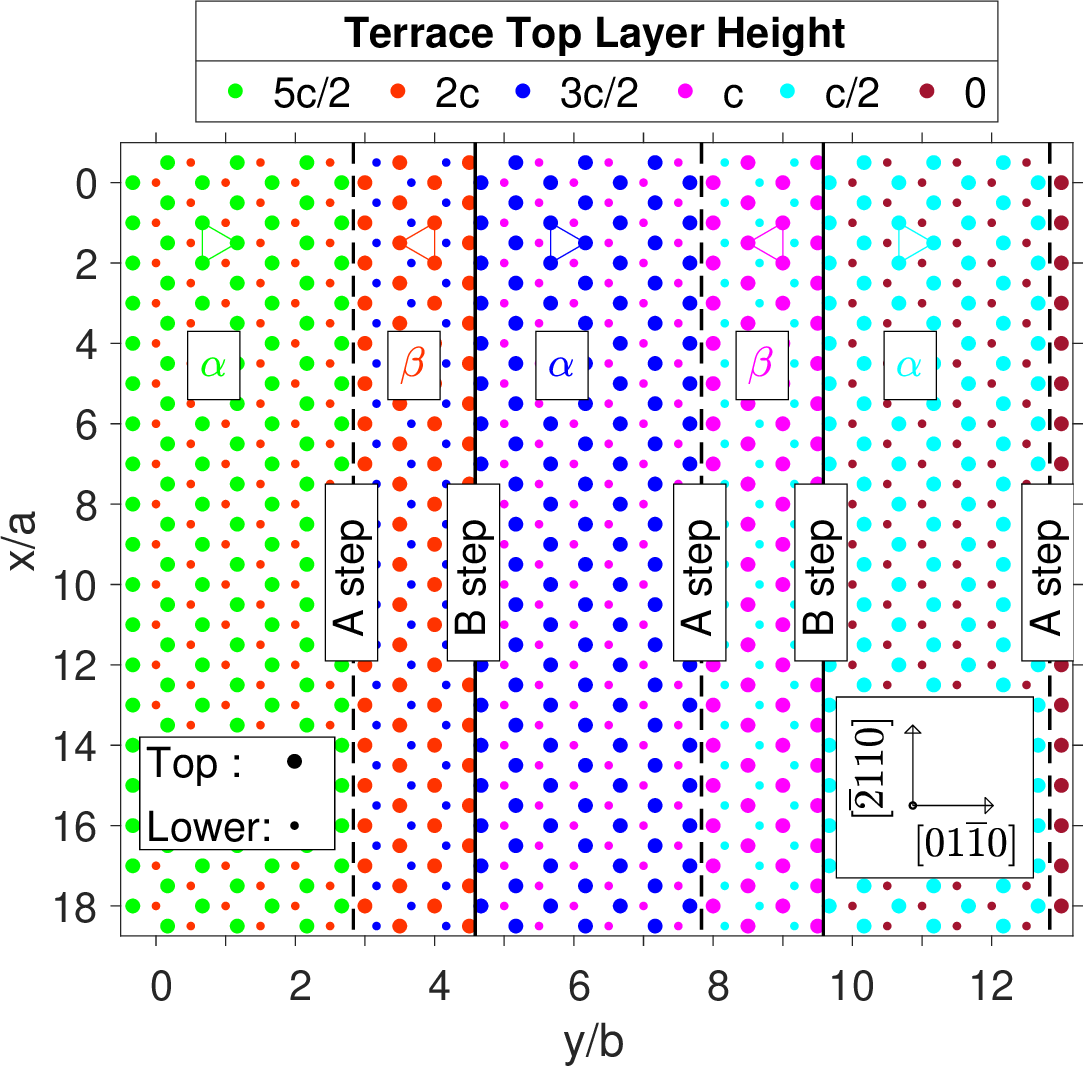}
\caption{Terrace and step structure of vicinal (0001) surface of an HCP-type crystal (see also Figs. 2 and 3 in \cite{2021_Ju_PRB_CTR}). Here we show a surface with an alpha terrace fraction $f_\alpha$ greater than 1/2. Large and small circles show in-plane positions of top-layer and second-layer atoms, respectively, with color indicating height. For GaN, only Ga atoms are shown, in unrelaxed (bulk) positions; not shown are N or passivating species of the surface reconstruction \cite{2021_Ju_PRB_CTR}. Orientation of triangle of top-layer atoms around $6_3$ screw axis shows difference between $\alpha$ and $\beta$ terraces. Atomic coordinates are given using orthohexagonal lattice parameters $a$, $b$, and $c$ \cite{1965_Otte_PhysStatSol9_441,2021_Ju_PRB_CTR}. Steps of height $c/2$ typically have lowest edge energy when they are normal to $[0 1 \overline{1} 0]$, $[1 0 \overline{1} 0]$, or $[1 \overline{1} 0 0]$. Steps in a sequence of a given azimuth have alternating structures, $A$ and $B$.  
\label{fig:A_B}}
\end{figure}

Most implementations of one-dimensional BCF theory presume that all steps have identical kinetic properties.
This is based on the assumption that steps have full-unit-cell heights, and thus identical structures owing to the crystal lattice periodicity.
However, when steps have fractional-unit-cell heights, the kinetic properties can differ from step to step.
This generally occurs for crystal symmetries which contain screw axes or glide planes, and can lead to fundamentally different growth behavior \cite{2004_vanEnckevort_ActaCrystA60_532}.
For example, on basal plane surfaces of crystals with hexagonal close-packed (HCP) or related structures (such as wurtzite GaN), which have a $6_3$ screw axis normal to the surface, it is common to find steps of half-unit-cell height because of the $\alpha \beta \alpha \beta$ stacking sequence of the lattice.
As shown in Fig.~\ref{fig:A_B}, on a vicinal surface the orientation of the atomic arrangements alternates between each $\alpha$ and $\beta$ layer, so that the structure and properties of the steps also alternate.
For such HCP-type systems, the adatom diffusivity is isotropic and equal on all terraces, and only the step properties alternate.
The lowest-energy steps are often normal to $\langle 0 1 \overline{1} 0 \rangle$ type directions, and the two resulting step structures are conventionally labelled $A$ and $B$ \cite{2001_Giesen_ProgSurfSci68_1,1999_Xie_PRL82_2749}.
(Face-centered cubic materials also have $A$ and $B$ type steps on close-packed $\{ 1 1 1 \}$ surfaces, but they do not alternate between successive terraces for a given step orientation \cite{2001_Giesen_ProgSurfSci68_1}.)
The kinetics of adatom attachment at $A$ and $B$ steps have been predicted to differ \cite{1999_Xie_PRL82_2749,2006_Xie_PRB_085314, 2011_Zaluska-Kotur_JAP109_023515,2010_Zaluska-Kotur_JNoncrystSolids356_1935,2013_Turski_JCrystGrowth367_115,2017_Xu_JChemPhys146_144702,2017_Chugh_ApplSurfSci422_1120,2020_Akiyama_JCrystGrowth532_125410,2020_Akiyama_JJAP59_SGGK03,ohka2020effect}, which can explain the alternating terrace widths and step morphologies often observed in HCP-type systems \cite{2006_Xie_PRB_085314,1979_Sunagawa_JCrystGrowth46_451,1982_vanderHoek_JCrystGrowth58_545,1999_Heying_JAP85_6470,2002_Chen_APL80_1358,2002_Zauner_JCrystGrowth240_14,2007_Krukowski_CrystResTechnol42_1281,2008_Zheng_PRB77_045303,2013_Lin_ApplPhysExp6_035503,2017_Pristovsek_physstatsolb254_1600711}.
Figure~\ref{fig:A_B} illustrates an example in which the $\alpha$ terraces are wider than the $\beta$ terraces.

\begin{figure}
\includegraphics[width=0.9\linewidth]{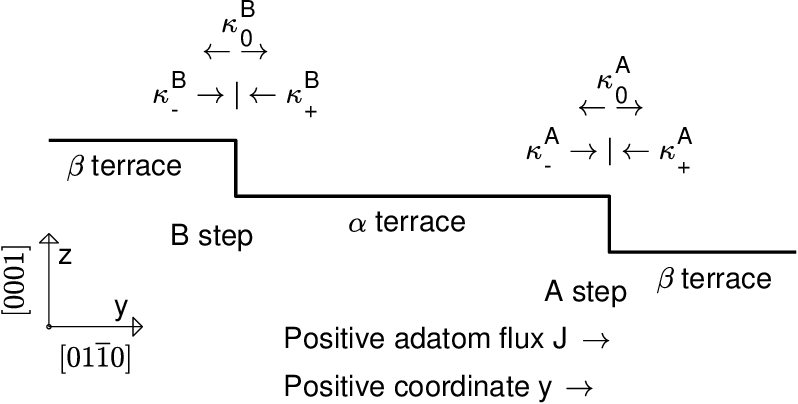}
\caption{Schematic of alternating terraces and steps for the BCF model. Vicinal $\{0001\}$ surfaces of HCP crystals have alternating $\alpha$ and $\beta$ terraces separated by $A$ and $B$ steps. Notations are indicated for the kinetic coefficients for adatom attachment from below $\kappa_+^j$ and above $\kappa_-^j$ and for adatom transmission $\kappa_0^j$. \label{fig:BCF}}
\end{figure}

Motivated by recent experimental results \cite{2020_Ju_NC} and surface X-ray scattering analysis \cite{2021_Ju_PRB_CTR}, here we develop a quantitative BCF model for surfaces with alternating step types.
We consider a simple one-dimensional model with an alternating sequence in the $y$ direction of two types of steps, $A$ and $B$, with properties that can differ, as shown in Figs.~\ref{fig:A_B} and \ref{fig:BCF}.
Related BCF models of alternating step or terrace properties have been developed previously \cite{2011_Zaluska-Kotur_JAP109_023515, 2010_Zaluska-Kotur_JNoncrystSolids356_1935,2007_Sato_EurPhysJB59_311,2006_Xie_PRB_085314,2005_Frisch_PRL94_226102,1991_Myers_APL59_2013}.
Discrete deposition-diffusion models with alternating step and terrace properties have also been presented \cite{ackerman2011boundary,zhao2015refined,2016_Zhao_PRB93_165411}.
Here we include the effects of step transparency (also known as step permeability, the transmission of adatoms across steps without incorporation) \cite{2003_Pierre-Louis_PRE68_021604,2003_Pierre-Louis_SurfSci529_114,2005_Krug_ISNM149_69,2007_Ranguelov_PRB75_245419} and step-step repulsion \cite{1999_Jeong_SurfSciRep34_171,2010_Patrone_PRE82_061601}.
We allow the kinetic and thermodynamic coefficients that determine the boundary conditions for adatom interaction with steps to differ for $A$ and $B$ steps, but assume that adatom deposition, diffusivity, and lifetime on $\alpha$ and $\beta$ terraces are identical.
We develop quasi-steady-state solutions for the adatom density distributions and the dynamics of the $\alpha$ and $\beta$ terrace fractions, and investigate how the full-steady-state terrace fraction depends upon growth rate and kinetic parameters.
Finally, we compare the BCF model predictions to recent \textit{in situ} microbeam X-ray scattering measurements of the terrace fraction during growth of GaN \cite{2020_Ju_NC}. 
In Appendix A, we connect a model for adatom diffusion between kinks on steps to the model for diffusion between steps on terraces, which gives relations between the kinetic coefficients involved in the step and kink boundary conditions, and provides a parameter that quantitatively characterizes step transparency.
In Appendix B, we consider cases with kinetics limited by diffusion, attachment, or a mixture on different terraces.
Our results are relevant to epitaxial growth of hexagonal wide-bandgap semiconductors such as GaN, AlN, and SiC, of current interest for opto-electronics \cite{2013_DenBaars_ActaMat61_945}, high-power electronics \cite{2018_Amano_JPhysD51_163001} and quantum information systems \cite{2017_Seo_PhysRevMat1_075002}.

\section{Burton-Cabrera-Frank theory for alternating step types}

In this section we develop a quasi-steady-state expression for the dynamics of the terrace fraction $f_\alpha$, and give an exact solution using matrices.
Examples of the full-steady-state adatom distributions and dynamics of $f_\alpha$ between such states are shown.
We then develop a simplified analytical solution, generally valid when the terrace widths are smaller than the adatom diffusion length.

\subsection{Exact quasi-steady-state solution}

Extending standard BCF theory \cite{1999_Jeong_SurfSciRep34_171,2003_Pierre-Louis_PRE68_021604,2005_Krug_ISNM149_69,2003_Pierre-Louis_SurfSci529_114,2004_Pierre-Louis_PRL93_165901} to a system with alternating types of terraces and steps, the continuity equation for the rate of change in the adatom density per unit area $\rho_i$ on terrace type $i = \alpha$ or $\beta$ is written as
\begin{equation}
    \frac{\partial \rho_i}{\partial t} = D \nabla_y^2 \rho_i - \frac{\rho_i}{\tau} + F,
    \label{eq:cont}
\end{equation}
where $D$ is the adatom diffusivity, $\tau$ is the adatom lifetime before evaporation, and $F$ is the deposition flux of adatoms per unit time and area. 
The four boundary conditions for the flux at the steps terminating opposite sides of each type of terrace can be written as
\begin{align}
    J_\alpha^+ &= - D \nabla_y \rho_\alpha^+
    = +\kappa_-^A (\rho_\alpha^+ - \rho_\mathrm{eq}^A)
    + \kappa_0^A (\rho_\alpha^+ - \rho_\beta^-), \label{eq:bc1} \\
    J_\alpha^- &= - D \nabla_y \rho_\alpha^-
    = -\kappa_+^B (\rho_\alpha^- - \rho_\mathrm{eq}^B)
    - \kappa_0^B (\rho_\alpha^- - \rho_\beta^+), \label{eq:bc2} \\
    J_\beta^+ &= - D \nabla_y \rho_\beta^+
    = +\kappa_-^B (\rho_\beta^+ - \rho_\mathrm{eq}^B)
    + \kappa_0^B (\rho_\beta^+ - \rho_\alpha^-), \label{eq:bc3} \\
    J_\beta^- &= - D \nabla_y \rho_\beta^-
    = -\kappa_+^A (\rho_\beta^- - \rho_\mathrm{eq}^A)
    - \kappa_0^A (\rho_\beta^- - \rho_\alpha^+), \label{eq:bc4}
\end{align}
where $J_i$ is the adatom surface flux on terrace $i$, $\kappa_+^j$ and $\kappa_-^j$ are the kinetic coefficients for adatom attachment at a step of type $j = A$ or $B$ from below or above, respectively,
$\kappa_0^j$ is the kinetic coefficient for transmission across the step,
and $\rho_\mathrm{eq}^j$ is the equilibrium adatom density at a step of type $j$.
A standard positive Ehrlich-Schwoebel (ES) barrier is given by $\kappa_+^j > \kappa_-^j$.
The $+$ or $-$ superscripts on $J_i$, $\rho_i$, and $\nabla_y \rho_i$ indicate evaluation at the terrace boundaries $y = + w_i/2$ or $y = - w_i/2$, respectively, where $w_i$ is the width of the terraces of type $i$ and the spatial coordinate $y$ is taken to be zero in the center of each terrace.
While there is a well-established convention for the definition of $A$ and $B$ steps owing to their different structures \cite{1999_Xie_PRL82_2749,2001_Giesen_ProgSurfSci68_1}, the definition of the $\alpha$ and $\beta$ terraces is somewhat arbitrary.
As shown in Figs.~\ref{fig:A_B} and~\ref{fig:BCF}, we adopt a convention in which the $\alpha$ terrace is above the $A$ step, and the $\beta$ terrace is above the $B$ step.

The last term in Eqs.~(\ref{eq:bc1}-\ref{eq:bc4}) accounts for step transparency \cite{2003_Pierre-Louis_PRE68_021604,2003_Pierre-Louis_SurfSci529_114,2005_Krug_ISNM149_69,2007_Ranguelov_PRB75_245419}, a phenomenon in which adatoms cross the step to exchange between neighboring terraces without attachment at a kink site on the step.
This process involves temporary adatom attachment to a step and some diffusion along the step, but with adatom detachment onto the opposite terrace before a kink is encountered.
Since the processes occurring along the step (in the $x$ direction) cannot be explicitly considered in this one-dimensional model for diffusion normal to the steps (in the $y$ direction), the transmission coefficients $\kappa_0^j$ are introduced to account for adatom densities attached to the steps that are not in equilibrium with the kinks. 
Appendix A gives a simple model of line diffusion of adatoms along a step between kinks that allows the kinetic coefficients $\kappa_+^j$, $\kappa_-^j$, and $\kappa_0^j$ to be related to the line diffusivity, kink attachment coefficients, and kink density.

The velocity $v_j$ of the $j$ type step can be obtained from the adatom fluxes arriving from each side, giving
\begin{align}
v_A &= \left ( J_\alpha^+ - J_\beta^- \right ) / \rho_0, \label{eq:va1} \\
v_B &= \left ( J_\beta^+ - J_\alpha^- \right ) / \rho_0,
\label{eq:vb1}
\end{align}
where $\rho_0$ is the density of lattice sites per unit area.

In both the continuity equation (\ref{eq:cont}) and the boundary conditions Eqs.~(\ref{eq:bc1}-\ref{eq:bc4}), we have neglected the ``advective'' terms due to the motion of the coordinate system and the boundaries with respect to the crystal lattice upon which the diffusion occurs.
Advection introduces a term $-v \rho_i$ into the adatom flux $J_i$ in addition to the diffusive term $-D \nabla_y \rho_i$, where $v$ is the velocity of the frame of reference of the flux relative to the lattice. 
This would contribute a term $(v_A + v_B) \nabla_y \rho_i / 2$ to the right-hand side of the continuity equation (\ref{eq:cont}) and terms $- v_j \rho_i^x$ to the left-hand sides of the boundary conditions Eqs.~(\ref{eq:bc1}-\ref{eq:bc4}), analogous to those used in one or both places in some previous work \cite{1988_Ghez_IBMJRD32_804,2003_Pierre-Louis_PRE68_021604,2003_Pierre-Louis_SurfSci529_114,2007_Dufay_PRB75_241304,2011_Zaluska-Kotur_JAP109_023515,2020_Guin_PRL124_036101}.
While the effects of these advective terms have been investigated
\cite{1988_Ghez_IBMJRD32_804,2007_Dufay_PRB75_241304,2020_Guin_PRL124_036101}, our neglect of them here is valid under the assumption that the adatom coverages are small, $\rho_i / \rho_0 << 1$.
We verify the self-consistency of neglecting advective terms in Supplemental Material \cite{BCF_supplemental}.

We assume that the adatom density profiles $\rho_i(y)$ have reached a quasi-steady-state where $\partial \rho_i / \partial t$ is negligible in the continuity equation, Eq.~(\ref{eq:cont}).
We still allow the terrace widths $w_i$ (and thus the $\rho_i$) to evolve relatively slowly with time.
The self-consistency of the quasi-steady-state approximation is analyzed in Supplemental Material \cite{BCF_supplemental}.
At quasi-steady-state, the general solution for the $\rho_i$ satisfying Eq.~(\ref{eq:cont}) with $\partial \rho_i / \partial t = 0$ is
\begin{equation}
    \rho_i = F \tau + C_{1i} \cosh \left ( \frac{y}{\sqrt{D \tau}} \right ) + C_{2i} \sinh \left ( \frac{y}{\sqrt{D \tau}} \right ),
    \label{eq:qss}
\end{equation}
where $C_{1i}$ and $C_{2i}$ are coefficients to be determined from the boundary conditions for each terrace type $i = \alpha$ or $\beta$.
The gradient $\nabla_y \rho_i$ that enters the boundary conditions is then
\begin{equation}
    \nabla_y \rho_i = \frac{C_{1i}}{\sqrt{D \tau}} \sinh \left ( \frac{y}{\sqrt{D \tau}} \right ) 
    + \frac{C_{2i}}{\sqrt{D \tau}} \cosh \left ( \frac{y}{\sqrt{D \tau}} \right ).
    \label{eq:grad}
\end{equation}
If we define the coefficients
\begin{equation}
c_i \equiv \cosh \left ( \frac{w_i}{2\sqrt{D \tau}} \right ),
\end{equation}
\begin{equation}
s_i \equiv \sinh \left ( \frac{w_i}{2\sqrt{D \tau}} \right ),
\end{equation}
for terrace types $i = \alpha$ and $\beta$, and dimensionless step kinetic parameters
\begin{align}
p_j &\equiv (\tau/D)^{1/2} \, \, \kappa_+^j, \\
q_j &\equiv (\tau/D)^{1/2} \, \, \kappa_-^j, \\
r_j &\equiv (\tau/D)^{1/2} \, \, \kappa_0^j,
\end{align}
for step types $j = A$ and $B$,
then we can use the quasi-steady-state solution Eq.~(\ref{eq:qss},\ref{eq:grad}) to write the boundary conditions Eq.~(\ref{eq:bc1}-\ref{eq:bc4}) as
\begin{equation}
    \mathcal{Q}\mathcal{C} = \mathcal{B}, \label{eq:MC}
\end{equation}
where $\mathcal{Q}$ is a matrix given by
\begin{widetext}
\begin{equation}
\mathcal{Q} =
\begin{bmatrix}
 + [ s_\alpha + (q_A + r_A) c_\alpha ] &
+ [ c_\alpha + (q_A + r_A) s_\alpha ] &
- r_A c_\beta & + r_A s_\beta \\
+ [ s_\alpha + (p_B + r_B) c_\alpha ] &
- [ c_\alpha + (p_B + r_B) s_\alpha ] &
- r_B c_\beta & - r_B s_\beta \\ 
- r_B c_\alpha & + r_B s_\alpha &
+ [ s_\beta + (q_B + r_B) c_\beta ] &
+ [ c_\beta + (q_B + r_B) s_\beta ] \\ 
- r_A c_\alpha & - r_A s_\alpha &
+ [ s_\beta + (p_A + r_A) c_\beta ] &
- [ c_\beta + (p_A + r_A) s_\beta ]
\end{bmatrix}
\label{eq:M}
\end{equation}
\end{widetext}
and the vectors $\mathcal{C}$ and $\mathcal{B}$ are given by
\begin{equation}
\mathcal{C} =
\begin{bmatrix}
C_{1\alpha} \\
C_{2\alpha} \\
C_{1\beta} \\
C_{2\beta}
\end{bmatrix},
\label{eq:C}
\end{equation}
\begin{equation}
\mathcal{B} =
\begin{bmatrix}
q_A (\rho_\mathrm{eq}^A - F \tau) \\
p_B (\rho_\mathrm{eq}^B - F \tau) \\ 
q_B (\rho_\mathrm{eq}^B - F \tau) \\ 
p_A (\rho_\mathrm{eq}^A - F \tau)
\end{bmatrix}.
\label{eq:B}
\end{equation}
The solution for the values of the four coefficients $C_{1i}$ and $C_{2i}$ of Eq.~(\ref{eq:qss}) is given by
\begin{equation}
    \mathcal{C} = \mathcal{Q}^{-1} \mathcal{B},
\label{eq:MB}
\end{equation}
where $\mathcal{Q}^{-1}$ is the inverse of $\mathcal{Q}$.
The quasi-steady-state step velocities can then be evaluated from expressions obtained using Eqs.~(\ref{eq:bc1}-\ref{eq:grad}),
\begin{align}
v_A &= - \sqrt{\frac{D}{\tau}} \left ( 
\frac{s_\alpha C_{1\alpha} + c_\alpha C_{2\alpha} 
+ s_\beta C_{1\beta} - c_\beta C_{2\beta}}
{\rho_0} \right ), \label{eq:va} \\
v_B &= - \sqrt{\frac{D}{\tau}} \left ( 
\frac{s_\alpha C_{1\alpha} - c_\alpha C_{2\alpha} 
+ s_\beta C_{1\beta} + c_\beta C_{2\beta}}
{\rho_0} \right ). \label{eq:vb}
\end{align}

The final expressions needed are those for the equilibrium adatom densities at the steps $\rho_\mathrm{eq}^j$ that enter the boundary conditions Eqs.~(\ref{eq:bc1}-\ref{eq:bc4}) and the vector $\mathcal{B}$.
These expressions include an effective repulsion between the steps owing to entropic and strain effects.
As in previous work \cite{1999_Jeong_SurfSciRep34_171,2010_Patrone_PRE82_061601}, we relate the equilibrium adatom density at a step to an adatom chemical potential $\mu_j$ via
\begin{equation}
    \rho_\mathrm{eq}^j = \rho_\mathrm{eq}^0 \exp(\mu_j/kT),
    \label{eq:rhoeqj}
\end{equation}
where $\rho_\mathrm{eq}^0 = (\rho_\mathrm{eq}^A \rho_\mathrm{eq}^B)^{1/2}$ is the mean equilibrium adatom density at zero growth rate, and $\mu_j$ depends on the terrace widths. 
In our case, with two different types of steps, $j = A$ and $B$, the chemical potentials are given by
\begin{equation}
    \frac{\mu_A}{kT} = -\frac{\mu_B}{kT} = M \equiv M_0 + \left ( \frac{\ell_\beta}{w_\beta} \right )^3 - \left ( \frac{\ell_\alpha}{w_\alpha} \right)^3. \label{eq:muAB}
\end{equation}
Here a non-zero offset $M_0$ arises from the difference in the adatom density in equilibrium with isolated $A$ or $B$ steps, and the $\ell_i$ are two step repulsion lengths, that can differ for the two types of terraces.
The step repulsion term prevents step collisions.
For example, if the alpha terrace width $w_\alpha$ approaches zero, the equilibrium adatom density for the $A$ step approaches zero and that for the $B$ step increases without limit.
One can see from Eqs.~(\ref{eq:bc1}-\ref{eq:vb1}) that this increases $J_\alpha^+$ and $J_\alpha^-$ and decreases $J_\beta^+$ and $J_\beta^-$, increasing $v_A$ and decreasing $v_B$, thus tending to increase $w_\alpha$.

\begin{figure}
\includegraphics[width=0.9\linewidth]{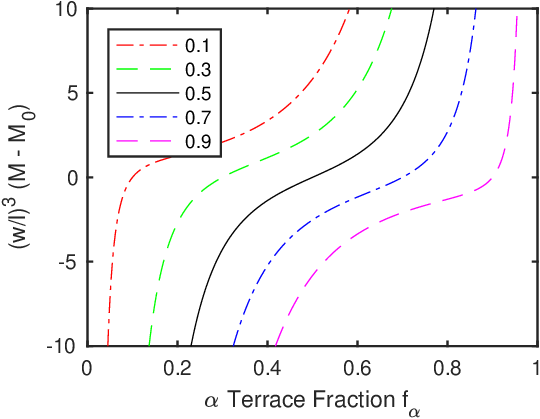}
\caption{Scaled and offset step chemical potential $(w/\ell)^3 (M - M_0)$ as a function of $f_\alpha$ from Eq.~(\ref{eq:Mfa}) for various values of $f_\alpha^0$ shown in legend. \label{fig:M0}}
\end{figure}

We consider the overall vicinal angle of the surface to fix the sum $w$ of the widths of $\alpha$ and $\beta$ terraces, so that the widths can be expressed as $w_i = f_i w$, where there is one independent terrace fraction $f_\alpha$, and the other is given by $f_\beta = 1 - f_\alpha$.
In this case we can express the chemical potentials using
\begin{equation}
     M(f_\alpha) = M_0 + \left ( \frac{\ell}{w} \right )^3 \left [ \left ( \frac{1 - f_\alpha^0}{1 - f_\alpha} \right )^3 - \left ( \frac{f_\alpha^0}{f_\alpha} \right)^3 \right ], \label{eq:Mfa}
\end{equation}
where the coefficients $\ell$ and $f_\alpha^0$ are related to the $\ell_i$ by
\begin{align}
    \ell_\alpha &= f_\alpha^0 \ell, \\
    \ell_\beta &= (1 - f_\alpha^0) \ell.
\end{align}
Here $\ell = \ell_\alpha + \ell_\beta$ is the total step repulsion length and the coefficient $f_\alpha^0 $ is the equilibrium terrace fraction at zero growth rate for $M_0 = 0$.
Figure~\ref{fig:M0} shows the dependence on $f_\alpha$ of the scaled step chemical potential $(w/\ell)^3 (M - M_0)$ for various values of $f_\alpha^0$.

\begin{figure}
\includegraphics[width=0.9\linewidth]{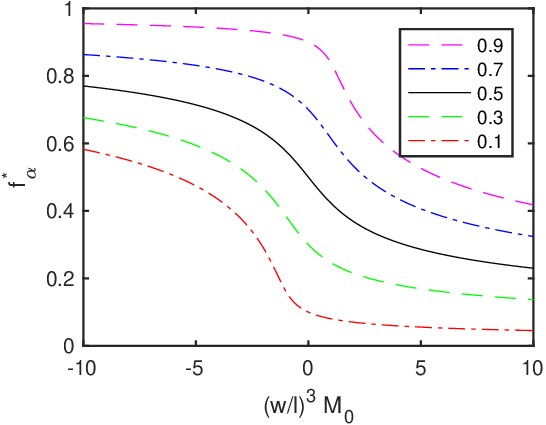}
\caption{Equilibrium terrace fraction at zero growth rate $f_\alpha^*$ as a function of the scaled step chemical potential offset $(w/\ell)^3 M_0$ for various values of $f_\alpha^0$ shown in legend. \label{fig:fsM}}
\end{figure}

\begin{figure}
\includegraphics[width=0.9\linewidth]{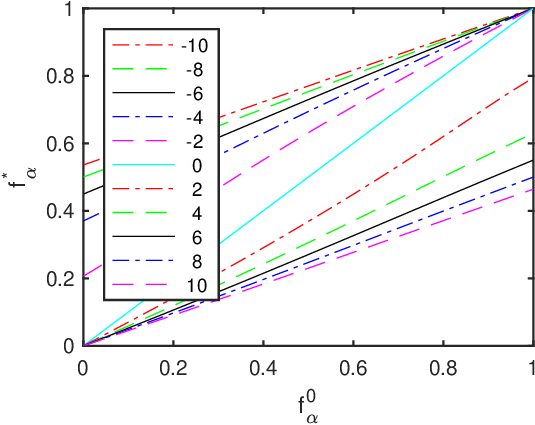}
\caption{Equilibrium terrace fraction at zero growth rate $f_\alpha^*$ as a function of $f_\alpha^0$ for various values of the scaled step chemical potential offset $(w/\ell)^3 M_0$ shown in legend. \label{fig:fsf}}
\end{figure}

For isolated steps, with $w \rightarrow \infty$, $A$ and $B$ steps can have a different equilibrium adatom densities, $\rho_\mathrm{eq}^A = \rho_\mathrm{eq}^0 \exp(M_0)$ and $\rho_\mathrm{eq}^B = \rho_\mathrm{eq}^0 \exp(-M_0)$, if $M_0$ is non-zero.
Recent \textit{ab initio} calculations \cite{2020_Akiyama_JCrystGrowth532_125410,2020_Akiyama_JJAP59_SGGK03,ohka2020effect} suggest that $A$ and $B$ steps can have different adatom attachment energies.
To estimate the offset $M_0$, one would have to consider not only adatom attachment energies at steps, but also the equilibrium concentration of adatoms attached to steps that result in zero net kink motion.
We discuss this in Appendix A.
When $\rho_\mathrm{eq}^A$ and $\rho_\mathrm{eq}^B$ differ, establishment of equilibrium on a vicinal surface with alternating step types requires that the step repulsion terms balance $M_0$ to give $M = 0$.
This occurs at a terrace fraction $f_\alpha = f_\alpha^*$, the equilibrium terrace fraction at zero growth rate, related to $M_0$ by the implicit expression
\begin{equation}
     -M_0 = \left ( \frac{\ell}{w} \right )^3 \left [ \left ( \frac{1 - f_\alpha^0}{1 - f_\alpha^*} \right )^3 - \left ( \frac{f_\alpha^0}{f_\alpha^*} \right)^3 \right ].  \label{eq:Mfs}
\end{equation}
Figure~\ref{fig:fsM} shows $f_\alpha^*$ as a function of the scaled offset $(w/\ell)^3M_0$ for various values of $f_\alpha^0$.
Inspection of Eqs.~(\ref{eq:Mfa}) and (\ref{eq:Mfs}) and Figs.~\ref{fig:M0} and \ref{fig:fsM} shows that the functional relationship between $f_\alpha^*$ and $-(w/\ell)^3 M_0$ is simply the inverse of the relationship between $(w/\ell)^3 (M -M_0)$ and $f_\alpha$.
Figure~\ref{fig:fsf} shows $f_\alpha^*$ as a function of $f_\alpha^0$ for various values of scaled $M_0$.
For $M_0 = 0$, one has simply $f_\alpha^* = f_\alpha^0$.

The net growth rate $G$ in monolayers per second (ML/s) is proportional to the sum of the step velocities,
\begin{equation}
 G = \frac{v_A + v_B}{w} = - \sqrt{\frac{D}{\tau}} \left ( 
\frac{2 s_\alpha C_{1\alpha} 
+ 2 s_\beta C_{1\beta}}
{w \rho_0} \right ).
\end{equation}
The rate of change of the $\alpha$ terrace fraction $f_\alpha$ is proportional to the step velocity difference,
\begin{equation}
 \frac{df_\alpha}{dt} = \frac{v_A - v_B}{w} = \sqrt{\frac{D}{\tau}} \left ( 
\frac{2 c_\beta C_{2\beta}
- 2 c_\alpha C_{2\alpha}}
{w \rho_0} \right ).
\label{eq:dfdt}
\end{equation}
This equation can be integrated to solve for the evolution of $f_\alpha(t)$ at quasi-steady-state.
To obtain the full-steady-state value of $f_\alpha$, the $A$ and $B$ step velocities must be equal and stable against fluctuations,
\begin{equation}
    \frac{df_\alpha}{dt} = 0,
    \label{eq:fullss}
\end{equation}
\begin{equation}
    \frac{\partial (df_\alpha/dt)}{\partial f_\alpha} < 0.
    \label{eq:stab}
\end{equation}
When the net growth rate is zero and the terrace fraction has reached its full-steady-state value, in this case equilibrium $f_\alpha = f_\alpha^*$, the step velocities are both zero, the diffusion fluxes are zero, the adatom densities are constant at a value $\rho_\alpha = \rho_\beta = \rho_\mathrm{eq}^A = \rho_\mathrm{eq}^B = \rho_\mathrm{eq}^0$, and the adatom chemical potentials at the steps are zero, $\mu_A = -\mu_B = 0$.

\subsection{Calculation of quasi-steady-state dynamics and full steady-state}

\begin{figure}
\includegraphics[width=0.8\linewidth]{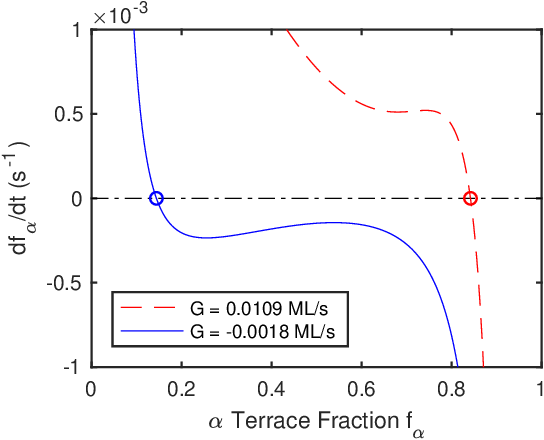}
\caption{Rate of change of the terrace fraction $df_\alpha/dt$ as a function of terrace fraction $f_\alpha$, calculated from Eq.~(\ref{eq:dfdt}) with parameter values given in Table~\ref{tab:table2}. The full-steady-state values $f_\alpha^\mathrm{ss}$ are marked with a circle.
\label{fig:dfdt1}}
\end{figure}

\begin{figure}
\includegraphics[width=0.8\linewidth]{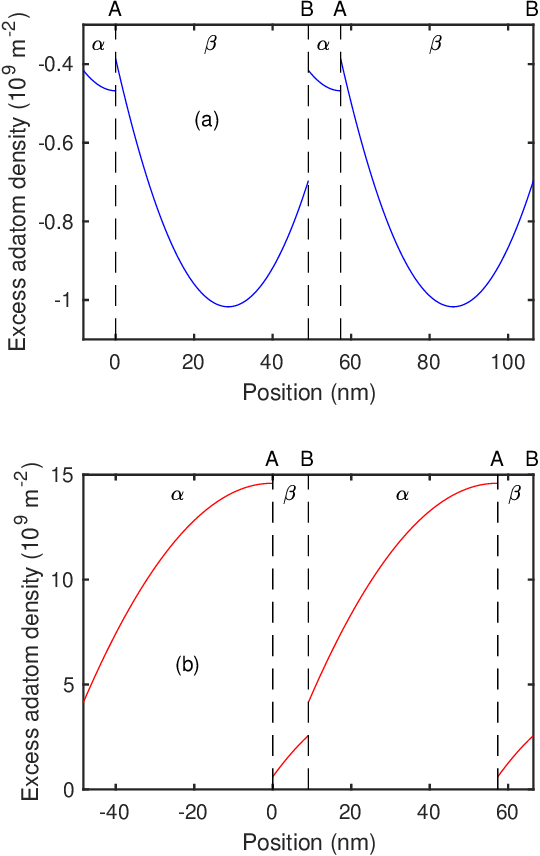}
\caption{Excess adatom density $\rho_i - \rho_\mathrm{eq}^0$ on a sequence of $\alpha$ and $\beta$ terraces corresponding to the full-steady-state solution, calculated with parameter values given in Table~\ref{tab:table2}, for (a) $F = 0$, $G = -0.0018$~ML/s, $f_\alpha^\mathrm{ss} = 0.146$, and (b) $F = 1.43 \times 10^{17}$~m$^{-2}$s$^{-1}$, $G = 0.0109$~ML/s, $f_\alpha^\mathrm{ss} = 0.837$. Origin of position coordinate is at an $A$ step. \label{fig:rhovy1}}
\end{figure}

Here we show some specific results calculated using the BCF theory for surfaces with alternating step types. Figure~\ref{fig:dfdt1} shows the quasi-steady-state rate of change of the terrace fraction $df_\alpha/dt$ as a function of terrace fraction $f_\alpha$, calculated from Eq.~(\ref{eq:dfdt}) with parameter values given in Table~\ref{tab:table2}.
These values are chosen to approximately match the experimental results for GaN $(0 0 0 1)$, using the fit SM1 in Section III below.
One curve is for a situation with no deposition flux, $F = 0$, where evaporation causes the net growth rate to be negative, $G = -0.0018$~ML/s, while the other is for a deposition flux of $F = 1.43\times10^{17}$~m$^{-2}$s$^{-1}$, giving a positive net growth rate of $G = 0.0109$~ML/s.
The full-steady-state values of terrace fraction $f_\alpha^\mathrm{ss}$ where $df_\alpha/dt = 0$ are marked in Fig.~\ref{fig:dfdt1} and given in Table~\ref{tab:table2}.
For these parameter values there is only a single full-steady-state solution for each curve, but from the non-monotonic shapes of the curves, one can see that two stable full-steady-state solutions can occur with other choices of parameter values.
(When the curve for $df_\alpha/dt$ crosses zero three times, only the outer two solutions with negative slope are stable; the middle solution with positive slope is unstable.)

\begin{table} 
\caption{ \label{tab:table2} Parameter values used in BCF theory calculations shown in Figs.~\ref{fig:dfdt1} - \ref{fig:fvst1}, from fit SM1 and estimates given below.
Also shown are derived values of $G$ and $f_\alpha^\mathrm{ss}$ for each $F$.}
\begin{ruledtabular}
\begin{tabular}{ l  l  l} 
$w = 5.73 \times 10^{-8}$ m & & $\rho_0 = 1.13 \times 10^{19}$ m$^{-2}$ \\
$\ell = 9.1 \times 10^{-10}$ m & & $\rho_\mathrm{eq}^0 = 3.44 \times 10^{12}$ m$^{-2}$ 	\\
$\tau = 1.66 \times 10^{-4}$ s & & $D = 1.35 \times 10^{-8}$ m$^2$ s$^{-1}$ \\
$\kappa_+^A = 1.0 \times 10^3$ m s$^{-1}$ & & $\kappa_+^B = 7.40 \times 10^{-1}$ m s$^{-1}$  \\
$\kappa_-^A = 1.0 \times 10^{-3}$ m s$^{-1}$ & & $\kappa_-^B = 1.0 \times 10^{-3}$ m s$^{-1}$  \\
$\kappa_0^A = 1.0 \times 10^{-3}$ m s$^{-1}$ & & $\kappa_0^B = 1.50 \times 10^0$ m s$^{-1}$  \\
$f_\alpha^0 = 0.441$ & & $M_0 = 0$ \\
\hline
\hline
Condition \# & 1 & 3 \\
\hline
$F$ $(10^{17}$ m$^{-2}$ s$^{-1}$) & $0$ & $1.43$ \\
$G$ (ML/s) & $-0.0018$ & $0.0109$ \\
$f_\alpha^\mathrm{ss}$ & $0.146$ & $.837$ \\
\end{tabular}
\end{ruledtabular}
\end{table}

\begin{figure}
\includegraphics[width=0.75\linewidth]{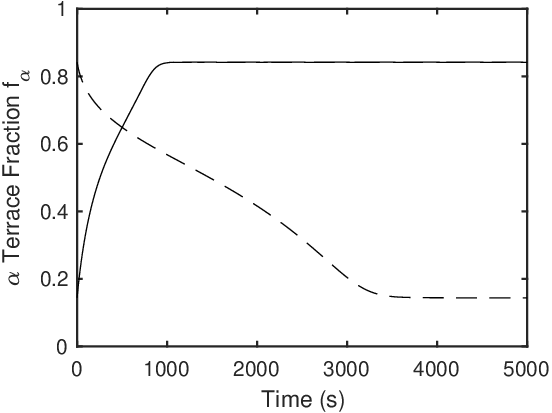}
\caption{Calculated time dependence of $f_\alpha$ obtained by integrating the quasi-steady-state result, Eq.~(\ref{eq:dfdt}), following changes between $G = -0.0018$ and $0.0109$~ML/s at $t = 0$. Solid and dashed curves are for increase or decrease of $G$, respectively. \label{fig:fvst1}}
\end{figure}

Figure~\ref{fig:rhovy1} shows the distribution of adatom density on a sequence of $\alpha$ and $\beta$ terraces at full steady-state, corresponding to the two growth rates shown in Fig.~\ref{fig:dfdt1}.
Since the fractional deviations from $\rho_\mathrm{eq}^0$ are very small, these are shown as the excess density $\rho_i - \rho_\mathrm{eq}^0$.
In Fig.~\ref{fig:rhovy1}(a), where $G$ is negative (i.e. evaporation is faster than deposition), the excess densities are negative and tend to go through minima on each terrace, while in Fig.~\ref{fig:rhovy1}(b), where $G$ is positive (i.e. deposition is faster than evaporation), the excess densities are positive and tend to go through maxima.
The discontinuities in $\rho_i$ at the steps reflect the differences in the adatom attachment coefficients from above and below, $\kappa_+^j$ and $\kappa_-^j$.
The low values of $\kappa_-^A$ and $\kappa_-^B$ used imply large ES barriers at the downhill (positive $y$) edges of the terraces, moving the maximum or minimum to that side.
The value of $\kappa_0^B$ gives significant transport across the $B$ step, reducing the difference in adatom densities across that step.

Figure~\ref{fig:fvst1} shows the calculated time dependence of $f_\alpha$ obtained by integrating the quasi-steady-state result, Eq.~(\ref{eq:dfdt}), for changes between the two conditions $G = -0.0018$~ML/s and $G = 0.0109$~ML/s.
Note that the predicted shapes are not simple exponentials.

\subsection{Analytical solution for non-transparent steps}

Because all four boundary conditions implied by Eq.~(\ref{eq:MC}) involve terms in all four coefficients $C_{1i}$ and $C_{2i}$, the explicit analytical solution of Eq.~(\ref{eq:MB}) for the coefficients gives very elaborate expressions. 
In the case of non-transparent steps, with $r_A = r_B = 0$, half of the elements of $\mathcal{Q}$ drop out and the boundary conditions split into two sets of two equations, each involving only two coefficients. 
In this case the analytical solutions are
\begin{widetext}
\begin{align}
C_{1\alpha} &= \frac{-F \tau [2 p_B q_A s_\alpha + (p_B + q_A)c_\alpha] 
+ (\rho_\mathrm{eq}^A + \rho_\mathrm{eq}^B) p_B q_A s_\alpha
+ (q_A \rho_\mathrm{eq}^A + p_B \rho_\mathrm{eq}^B) c_\alpha}
{(p_B + q_A)(s_\alpha^2 + c_\alpha^2)
+ 2(1 + p_B q_A) s_\alpha c_\alpha}, \label{eq:Ci1} \\
C_{2\alpha} &= \frac{F \tau (p_B - q_A) s_\alpha
+ (\rho_\mathrm{eq}^A - \rho_\mathrm{eq}^B) p_B q_A c_\alpha
+ (q_A \rho_\mathrm{eq}^A - p_B \rho_\mathrm{eq}^B) s_\alpha}
{(p_B + q_A)(s_\alpha^2 + c_\alpha^2)
+ 2(1 + p_B q_A) s_\alpha c_\alpha}, \label{eq:Ci2} \\
C_{1\beta} &= \frac{-F \tau [2 p_A q_B s_\beta + (p_A + q_B)c_\beta] 
+ (\rho_\mathrm{eq}^B + \rho_\mathrm{eq}^A) p_A q_B s_\beta
+ (q_B \rho_\mathrm{eq}^B + p_A \rho_\mathrm{eq}^A) c_\beta}
{(p_A + q_B)(s_\beta^2 + c_\beta^2)
+ 2(1 + p_A q_B) s_\beta c_\beta}, \label{eq:Ci3} \\
C_{2\beta} &= \frac{F \tau (p_A - q_B) s_\beta
+ (\rho_\mathrm{eq}^B - \rho_\mathrm{eq}^A) p_A q_B c_\beta
+ (q_B \rho_\mathrm{eq}^B - p_A \rho_\mathrm{eq}^A) s_\beta}
{(p_A + q_B)(s_\beta^2 + c_\beta^2)
+ 2(1 + p_A q_B) s_\beta c_\beta}. \label{eq:Ci4} 
\end{align}
\end{widetext}

\subsection{Analytical solution for transparent steps}

To obtain an analytical solution of Eq.~(\ref{eq:MB}) including the effects of step transparency, we can work with an alternative, mathematically equivalent formulation of the boundary conditions \cite{2003_Pierre-Louis_PRE68_021604}
\begin{align}
    J_\alpha^+ &= - D \nabla_y \rho_\alpha^+
    = +\kapt_-^A (\rho_\alpha^+ - \rhot_\mathrm{eq}^A), \label{eq:bca1} \\
    J_\alpha^- &= - D \nabla_y \rho_\alpha^-
    = -\kapt_+^B (\rho_\alpha^- - \rhot_\mathrm{eq}^B), \label{eq:bca2} \\
    J_\beta^+ &= - D \nabla_y \rho_\beta^+
    = +\kapt_-^B (\rho_\beta^+ - \rhot_\mathrm{eq}^B), \label{eq:bca3} \\
    J_\beta^- &= - D \nabla_y \rho_\beta^-
    = -\kapt_+^A (\rho_\beta^- - \rhot_\mathrm{eq}^A), \label{eq:bca4}
\end{align}
where the quantities with tildes are defined as
\begin{align}
\kapt_+^j &\equiv \frac{\kapsq^j}{\kappa_-^j}, \label{eq:kt1} \\
\kapt_-^j &\equiv \frac{\kapsq^j}{\kappa_+^j}, \label{eq:kt2} \\
\rhot_\mathrm{eq}^j &\equiv \rho_\mathrm{eq}^j + \frac{v_j \rho_0 \kappa_0^j}{\kapsq^j}, \label{eq:kt3}
\end{align}
using a sum of products of kinetic coefficients for the type $j = \alpha$ or $\beta$ step denoted as
\begin{equation}
\kapsq^j \equiv \kappa_+^j \kappa_-^j + \kappa_+^j \kappa_0^j + \kappa_-^j \kappa_0^j. \label{eq:kt4}
\end{equation}
Note that in Eq.~(\ref{eq:kt3}) the effective equilibrium adatom density $\rhot_\mathrm{eq}^j$ at a step of type $j$ depends on the step velocity $v_j$.
The physical significance of $\kapt_+^j$, $\kapt_-^j$, and $\rhot_\mathrm{eq}^j$ are discussed in Appendix A.

The boundary conditions can be written as
\begin{equation}
    \tilde{\mathcal{Q}} \mathcal{C} = \tilde{\mathcal{B}}, \label{eq:MCt}
\end{equation}
where $\tilde{\mathcal{Q}}$ and $\tilde{\mathcal{B}}$ are given by
\begin{align}
&\tilde{\mathcal{Q}} = \nonumber \\
&\begin{bmatrix}
s_\alpha + \qt_A c_\alpha &
c_\alpha + \qt_A s_\alpha &
0 & 0 \\
s_\alpha + \pt_B c_\alpha &
- c_\alpha - \pt_B s_\alpha &
0 & 0 \\ 
0 & 0 &
s_\beta + \qt_B c_\beta &
c_\beta + \qt_B s_\beta \\ 
0 & 0 &
s_\beta + \pt_A c_\beta &
- c_\beta - \pt_A s_\beta
\end{bmatrix},
\label{eq:Mt}
\end{align}
\begin{equation}
\tilde{\mathcal{B}} =
\begin{bmatrix}
\qt_A (\rhot_\mathrm{eq}^A - F \tau) \\
\pt_B (\rhot_\mathrm{eq}^B - F \tau) \\ 
\qt_B (\rhot_\mathrm{eq}^B - F \tau) \\ 
\pt_A (\rhot_\mathrm{eq}^A - F \tau)
\end{bmatrix},
\label{eq:Bt}
\end{equation}
using alternative dimensionless step kinetic parameters
\begin{align}
\pt_j &\equiv \sqrt{\frac{\tau}{D}} \kapt_+^j = \frac{p_j q_j + p_j r_j + q_j r_j}{q_j}, \\
\qt_j &\equiv \sqrt{\frac{\tau}{D}} \kapt_-^j = \frac{p_j q_j + p_j r_j + q_j r_j}{p_j},
\label{eq:ptqt}
\end{align}
for step types $j = A$ and $B$.
As in the case of non-transparent steps, these boundary conditions consist of two sets of two equations, each involving only two coefficients, $C_{1i}$ and $C_{2i}$ with $i = \alpha$ or $\beta$. 
The solutions are the same as Eqs.~(\ref{eq:Ci1}-\ref{eq:Ci4}), with $p_j$, $q_j$, and $\rho_\mathrm{eq}^j$ replaced by $\pt_j$, $\qt_j$, and $\rhot_\mathrm{eq}^j$, respectively.
Unfortunately, since the $\rhot_\mathrm{eq}^j$ that appear in the $C_{1i}$ and $C_{2i}$ depend upon the step velocities $v_j$, which in turn depend upon the $C_{1i}$ and $C_{2i}$ via Eqs.~(\ref{eq:va}-\ref{eq:vb}), this still does not provide an explicit solution for the $C_{1i}$ and $C_{2i}$.

\subsection{Simplified analytical solution for small terrace width}

It is very useful to consider some broadly applicable limits that simplify the analytical solution, allowing the full-steady-state terrace fraction and its quasi-steady-state dynamics to be expressed in terms of the net growth rate.
We start with Eqs.~(\ref{eq:Ci1}-\ref{eq:Ci4}), with $p_j$, $q_j$, and $\rho_\mathrm{eq}^j$ replaced by $\pt_j$, $\qt_j$, and $\rhot_\mathrm{eq}^j$, respectively.
In the limit where the diffusion length within an adatom lifetime is much larger than the terrace widths, $\sqrt{D \tau} >> w$, the adatom distributions $\rho_i(y)$ are quadratic in $y$, their gradients $\nabla_y \rho_i(y)$ are linear in $y$, and the Laplacians $\nabla_y^2 \rho_i$ are constant. 
In Eqs.~(\ref{eq:Ci1}-\ref{eq:Ci4}), the coefficients $c_i$ can be set equal to unity, and the coefficients $s_i$ are small quantities given by $s_i = w_i/(2 \sqrt{D \tau})$.
In the limits $M_0 << 1$ and $\ell_i << w_i$, we have $M << 1$ and $\exp(M) \approx 1 + M$, so that the adatom densities $\rho_i$ do not differ much from $\rho_\mathrm{eq}^0$, and thus the adatom evaporation flux is relatively uniform at $\rho_\mathrm{eq}^0/\tau$.
Assuming the second term in Eq.~(\ref{eq:kt3}) is small,
we can replace $\rhot_\mathrm{eq}^A$ and $\rhot_\mathrm{eq}^B$ by $\rho_\mathrm{eq}^0$, except in the difference $(\rhot_\mathrm{eq}^A-\rhot_\mathrm{eq}^B)$.
The formulas for $C_{1i}$ simplify to be 
\begin{equation}
C_{1\alpha} \approx C_{1\beta} \approx \rho_\mathrm{eq}^0 - F \tau.  \label{eq:Ca1}
\end{equation}
The net growth rate is then simply given by
\begin{equation}
    G \approx \frac{F - \rho_\mathrm{eq}^0 / \tau}{\rho_0},  \label{eq:Ga}
\end{equation}
which is the difference between the deposition flux $F$ and a uniform evaporation flux $\rho_\mathrm{eq}^0 / \tau$, converted to ML/s using $\rho_0$.

If we also assume that the parameters $\pt_j$ and $\qt_j$ are generally greater than unity owing to large adatom lifetimes $\tau$, so that $\pt_A \qt_B >> 1$ and $\pt_B \qt_A >> 1$, we can write the expressions for the $C_{2i}$ as
\begin{align}
C_{2\alpha} &\approx \frac{\sqrt{D \tau}}{w} \big [ R_\alpha (\rhot_\mathrm{eq}^A - \rhot_\mathrm{eq}^B) + S_\alpha \rho_0 G \big ], \label{eq:Ca2a} \\
C_{2\beta} &\approx \frac{\sqrt{D \tau}}{w} \big [ R_\beta (\rhot_\mathrm{eq}^B - \rhot_\mathrm{eq}^A) + S_\beta \rho_0 G \big ], \label{eq:Ca3a}
\end{align}
where each contains a term that is proportional to the net growth rate $G$.
The coefficients are given by
\begin{align}
R_\alpha &\equiv \frac{w}{D} \left ( \frac{\kappa_+^A}{\kapsq^A} + \frac{\kappa_-^B}{\kapsq^B} + \frac{w f_\alpha}{D} \right )^{-1},  \label{eq:Ra} \\
R_\beta &\equiv  \frac{w}{D} \left ( \frac{\kappa_+^B}{\kapsq^B} + \frac{\kappa_-^A}{\kapsq^A} + \frac{w (1 - f_\alpha)}{D} \right )^{-1},  \label{eq:Rb} \\
S_\alpha &\equiv \frac{R_\alpha w f_\alpha}{2} \left ( \frac{\kappa_+^A}{\kapsq^A} - \frac{\kappa_-^B}{\kapsq^B} \right ),  \label{eq:Sa} \\
S_\beta &\equiv  \frac{R_\beta w (1 - f_\alpha)}{2} \left ( \frac{\kappa_+^B}{\kapsq^B} - \frac{\kappa_-^A}{\kapsq^A} \right ),  \label{eq:Sb}
\end{align}
where the $R_i$ are positive and dimensionless and the $S_i$ have dimensions of time.
The step velocities of Eqs.~(\ref{eq:va}-\ref{eq:vb}) become
\begin{align}
    v_A &= \frac{w G}{2} + \frac{D}{\rho_0 w} \big [ (R_\alpha + R_\beta) (\rhot_\mathrm{eq}^B - \rhot_\mathrm{eq}^A) + (S_\beta - S_\alpha) \rho_0 G \big ],  \label{eq:vas}\\
    v_B &= \frac{w G}{2} + \frac{D}{\rho_0 w} \big [ (R_\alpha + R_\beta) (\rhot_\mathrm{eq}^A - \rhot_\mathrm{eq}^B) + (S_\alpha - S_\beta) \rho_0 G \big ]. \label{eq:vbs}
\end{align}
The difference of the effective equilibrium step adatom densities also contains a term that is proportional to $G$,
\begin{equation}
    \rhot_\mathrm{eq}^A - \rhot_\mathrm{eq}^B = \frac{2\rho_\mathrm{eq}^0 M + \rho_0 G \big [ S_0 + R_0 (S_\beta - S_\alpha) \big ]}{1 + R_0 (R_\alpha + R_\beta)}, \label{eq:rhoteq}
\end{equation}
where the new coefficients are given by
\begin{align}
    R_0 &\equiv \frac{D}{w} \left ( \frac{\kappa_0^A}{\kapsq^A} + \frac{\kappa_0^B}{\kapsq^B} \right ), \label{eq:R0} \\
    S_0 &\equiv \frac{w}{2} \left ( \frac{\kappa_0^A}{\kapsq^A} - \frac{\kappa_0^B}{\kapsq^B} \right ).  \label{eq:S0}
\end{align}
The rate of change of $f_\alpha$ becomes
\begin{equation}
    \frac{df_\alpha}{dt} = K^\mathrm{dyn}(f_\alpha) \left ( \frac{G}{K^\mathrm{ss}(f_\alpha)} - \frac{4 M(f_\alpha) \rho_\mathrm{eq}^0}{w \rho_0} \right ), \label{eq:dfdta}
\end{equation}
where we have introduced the combined kinetic coefficient functions $K^\mathrm{ss}(f_\alpha)$ and $K^\mathrm{dyn}(f_\alpha)$, defined by
\begin{align}
    K^\mathrm{ss}(f_\alpha) &\equiv \frac{w}{2 \big [ -S_0 + (S_\beta - S_\alpha)/(R_\alpha + R_\beta) \big ]},
    \label{eq:Dk} \\
    K^\mathrm{dyn}(f_\alpha) &\equiv \frac{D}{w [ R_0 + 1/(R_\alpha + R_\beta)]}.
    \label{eq:KD}
\end{align}
These functions have the same dimensions as the individual $\kappa_x^j$ coefficients (length/time). 
$K^\mathrm{dyn}(f_\alpha)$ is always positive; $K^\mathrm{ss}(f_\alpha)$ depends on the differences in the $\kappa_x^j$, such that in the limit where all $\kappa_x^j$ are equal, $K^\mathrm{ss} \rightarrow \infty$.
In this case the influence of $G$ on $f_\alpha$ becomes negligible, and the full-steady-state $\alpha$ terrace fraction is always $f_\alpha^\mathrm{ss} = f_\alpha^*$ (i.e. the value where $M = 0$), independent of $G$. 

The general equation to obtain the full steady-state is
\begin{equation}
    G^\mathrm{ss}(f_\alpha) = \frac{4 \, K^\mathrm{ss}(f_\alpha) \, M(f_\alpha) \rho_\mathrm{eq}^0}{w \rho_0}.
    \label{eq:fSSG}
\end{equation}
This equation for $G^\mathrm{ss}(f_\alpha)$ can be inverted to obtain a master curve for the full-steady-state value $f_\alpha^\mathrm{ss}$ as a function of $G$.
For both the dynamics Eq.~(\ref{eq:dfdta}) and the full steady-state Eq.~(\ref{eq:fSSG}),
the six step attachment parameters enter through the six combinations in the coefficients $R_i$, $S_i$, $R_0$, and $S_0$.
The adatom diffusivity $D$ enters only in the ratios $D/\kappa_x^j$ and the product $DM$.
The only dependence on $\tau$ and $F$ is through their combination into $G$, Eq.~(\ref{eq:Ga}).

The curve $G^\mathrm{ss}(f_\alpha)$ always passes through $G = 0$ at $f_\alpha = f_\alpha^*$, since $M$ is zero there.
The slope of the curve at $f_\alpha = f_\alpha^*$ is given by
\begin{equation}
    G^* \equiv \left . \frac{dG^\mathrm{ss}}{df_\alpha} \right |_{f_\alpha^*} = \frac{4 \, K^\mathrm{ss}(f_\alpha^*) M^\prime(f_\alpha^*) \rho_\mathrm{eq}^0}{w \rho_0}, \label{eq:Gstar} 
\end{equation}
where $M^\prime$ is the derivative
\begin{equation}
    M^\prime(f_\alpha) \equiv \frac{dM}{df_\alpha}
    = \frac{3 \ell^3}{w^3} \left [ \frac{(1 - f_\alpha^0)^3}{(1 - f_\alpha)^4} + \frac{(f_\alpha^0)^3}{(f_\alpha)^4} \right ].
    \label{eq:Mprime}
\end{equation}
The sign of the slope of $G^\mathrm{ss}(f_\alpha)$, and thus $f_\alpha^\mathrm{ss}(G)$, is determined by the sign of $K^\mathrm{ss}(f_\alpha^*)$, since all other factors are positive.

One can write Eq.~(\ref{eq:dfdta}) as
\begin{equation}
    \frac{df_\alpha}{dt} = \frac{K^\mathrm{dyn}(f_\alpha)}{K^\mathrm{ss}(f_\alpha)} \big [ G - G^\mathrm{ss}(f_\alpha) \big ]. \label{eq:dfdta2a}
\end{equation}
This form makes it clear that, near $f_\alpha^*$, $f_\alpha$ is always stable to a small perturbation from its full-steady-state value, $\Delta f_\alpha \equiv f_\alpha - f_\alpha^\mathrm{ss}(G)$.
For example, when $K^\mathrm{ss}$ is positive, and $\Delta f_\alpha$ is positive, then $G - G^\mathrm{ss}(f_\alpha) \approx - G^* \Delta f_\alpha$ will be negative, and the perturbation will decay.
The relaxation time $t^*$ of the perturbation can be obtained by substituting this approximation into Eq.~(\ref{eq:dfdta2a}) to give
\begin{equation}
    \frac{1}{t^*} \equiv \frac{-1}{\Delta f_\alpha} \frac{df_\alpha}{dt} \approx \frac{K^\mathrm{dyn}(f_\alpha^*) \, G^*}{K^\mathrm{ss}(f_\alpha^*)}
    = \frac{4 \, K^\mathrm{dyn}(f_\alpha^*) M^\prime(f_\alpha^*) \rho_\mathrm{eq}^0}{w \rho_0}. \label{eq:tstar}
\end{equation}

Away from $f_\alpha = f_\alpha^*$, the solutions can become unstable.
The stability criterion Eq.~(\ref{eq:stab}) can be written as
\begin{equation}
    \left . \frac{\partial (df_\alpha/dt)}{\partial f_\alpha} \right |_{f_\alpha^\mathrm{ss}} 
    = - \frac{K^\mathrm{dyn}(f_\alpha^\mathrm{ss})}{K^\mathrm{ss}(f_\alpha^\mathrm{ss})}
    \left . \frac{\partial G^\mathrm{ss}}{\partial f_\alpha} \right |_{f_\alpha^\mathrm{ss}}
    < 0.
    \label{eq:stab2}
\end{equation}
Thus the full-steady-state solution is stable whenever the slope of $G^\mathrm{ss}(f_\alpha)$ has the same sign as $K^\mathrm{ss}(f_\alpha)$.

Criteria on parameter values for the self-consistency of the approximations used to obtain the simplified analytical solution are given in Supplemental Material \cite{BCF_supplemental}. For the parameter ranges we consider, these criteria are generally satisfied, confirming the validity of this solution.
We have also checked that the exact solution obtained using the matrix equations Eqs.~(\ref{eq:MC}-\ref{eq:MB}) agrees with the simplified analytical solution when the criteria are satisfied.

In the general model, e.g. Eqs.~(\ref{eq:cont})-(\ref{eq:vb1}) and Eqs.~(\ref{eq:rhoeqj})-(\ref{eq:Mfa}), there are 15 fundamental variables ($F$, $\tau$, $\rho_0$, $w$, $D$, $\rho_\mathrm{eq}^0$, $M_0$, $\ell$, $f_\alpha^0$, and the six $\kappa_x^j$).
In the simplified analytical solution developed in this section, Eqs.~(\ref{eq:dfdta}-\ref{eq:fSSG}), 12 of the variables enter only through 9 combinations  ($G = [F - \rho_\mathrm{eq}^0/\tau]/\rho_0$, $D \rho_\mathrm{eq}^0 M_0$, $D \rho_\mathrm{eq}^0 \ell^3$, and the six ratios $D/\kappa_x^j$), leaving 12 independent variables that determine the behavior.
The ratios $D/\kappa_x^j$ have been named ``kinetic lengths'' \cite{2003_Pierre-Louis_SurfSci529_114,2004_Pierre-Louis_PRL93_165901,2005_Krug_ISNM149_69,2010_Patrone_PRE82_061601}.
Kinetic lengths much smaller or larger than the terrace widths typically give diffusion- or attachment-limited kinetics, respectively.
Appendix B shows how the expressions developed above for the simplified analytical solution reduce to simpler expressions for cases in which the adatom kinetics on the terraces are limited by diffusion or by attachment/detachment at steps.

\section{Comparison of BCF theory to X-ray measurements during OMVPE}

A primary motivation for the above development of BCF theory for surfaces with alternating step types has been to compare predictions with recent experimental measurements during step-flow growth and evaporation of GaN $(0 0 0 1)$ at $T = 1073$~K under organo-metallic vapor phase epitaxy (OMVPE) conditions \cite{2020_Ju_NC}.
These microbeam surface X-ray scattering measurements determined the steady-state terrace fraction $f_\alpha^\mathrm{ss}$ as a function of growth conditions, as well as typical time constants for the dynamics of $f_\alpha(t)$ upon changing conditions.
The measured values are summarized in Table~\ref{tab:tableM}, along with theory fit values described below.

While we do not explicitly model the potentially complex surface chemistry of OMVPE in this work, we expect that the basic framework of BCF theory can be applied, with the chemical states of the adatoms, steps, and terraces affecting the parameter values in the model.
The observed GaN growth rate \cite{2020_Ju_NC} has a simple transport-limited behavior, with a deposition flux $F$ that is linearly proportional to the supply of Ga precursor, since the N precursor is supplied in excess.
Under the conditions studied, the proportionality is independent of temperature, indicating that precursor reactions are not rate-limiting, as has been considered in some previous BCF models \cite{2000_Pimpinelli_SurfSci445_L23,2008_Chua_APL92_013117,2019_Hanada_PhysRevMater3_103404, 2020_Redkov_CGD20_2590}.
For each of the two deposition fluxes used ($F = 0$ and $1.43 \times 10^{17}$~m$^{-2}$~s$^{-1}$),
carrier gas compositions with and without H$_2$ were employed. 
The addition of H$_2$ to the carrier gas reduces the adatom lifetime $\tau$ and increases the evaporation flux $\rho_\mathrm{eq}^0 / \tau$ from negligible to $2.0 \times 10^{16}$~m$^{-2}$~s$^{-1}$.
From Eq.~(\ref{eq:Ga}), one can see that this slightly reduces the net growth rate $G$, which is proportional to the difference between the deposition and evaporation fluxes; at zero deposition flux, $G$ is negative.

To apply the BCF model to the GaN OMVPE environment, we have to consider terrace, step, and adatom structures that are more complex than in simple cases such as vacuum deposition of elemental metals.
The GaN terraces have a surface reconstruction involving passivation by adsorbed species such as H \cite{2020_Ju_NC,2021_Ju_PRB_CTR}.
The mobile ``adatoms" likely involve both Ga and N species.
We expect the chemistry of the environment to affect the kinetics of their diffusion and attachment at $A$ and $B$ steps, as found in previous studies of epitaxial growth in chemically active environments \cite{1998_Kalff_PRL81_1255,2009_Yin_APL94_183107,2017_Pristovsek_physstatsolb254_1600711,2020_Akiyama_JCrystGrowth532_125410,2020_Akiyama_JJAP59_SGGK03,ohka2020effect}.

As shown in Table~\ref{tab:tableM}, the experiments \cite{2020_Ju_NC} give a monotonic increase of $f_\alpha^\mathrm{ss}$ with $G$, and characteristic times for relaxation of $f_\alpha(t)$ upon changing conditions.
Our BCF model predicts the dependence of the full-steady-state terrace fraction on growth rate $f_\alpha^\mathrm{ss}(G)$, as well as the dynamics of the transitions when $G$ is changed.
We can compare calculated values to these measurements to understand the implications for the physics in the model, such as the differences between adatom attachment kinetics at $A$ and $B$ steps.

We have previously presented fits \cite{2020_Ju_NC} of a version of our BCF theory with $M_0$ fixed to zero, to the experimental results using only a single relaxation time $t_\mathrm{rel}$ for each of the transitions, where $t_\mathrm{rel}$ is the time for $\Delta f_\alpha(t) / \Delta f_\alpha(0)$ to reach $1/e = 37\%$.
Because the predicted relaxation of $f_\alpha(t)$ can be significantly non-exponential, as shown in Fig.~\ref{fig:fvst1}, here we have fit the theory to three measured characteristic times for each transition, rather than just a single relaxation time $t_\mathrm{rel}$.
The three times given in Table~\ref{tab:tableM}, $t_{80}$, $t_{50}$, and $t_{20}$, are the times for the normalized deviation of the terrace fraction from its steady-state value, $\Delta f_\alpha(t) / \Delta f_\alpha(0)$, to reach 80\%, 50\%, and 20\%, respectively, after a change of growth rate at $t = 0$.
Details of the extraction of $t_{80}$, $t_{50}$, and $t_{20}$ from the experimental data are given in Supplemental Material \cite{BCF_supplemental}.
Here we also allow $M_0$ to deviate from zero in the fits of BCF theory.

\begin{table} 
\caption{ \label{tab:tableM} Comparison of measured values (left columns) and calculated values from the four best fits of the simplified analytical BCF model.}
\begin{ruledtabular}
\begin{tabular}{ c | c || c || c | c | c | c}
\multicolumn {2} {c ||} {Condition} & \multicolumn {5} {c} {Steady-State Terrace Fraction $f_\alpha^\mathrm{ss}$} \\
\hline
\# & $G$ (ML/s) & Measured & SD3 & SM1 & SM2 & SM3 \\
\hline
$1$ & $-0.0018$ & $0.111 \pm 0.013$ & $0.117$ & $0.146$ & $0.153$ & $0.145$ \\
$2$ & $0.0000$  & $0.461 \pm 0.018$ & $0.464$ & $0.441$ & $0.461$ & $0.442$ \\
$3$ & $0.0109$  & $0.811 \pm 0.014$ & $0.828$ & $0.837$ & $0.843$ & $0.832$ \\
$4$ & $0.0127$  & $0.868 \pm 0.011$ & $0.839$ & $0.848$ & $0.853$ & $0.843$ \\
\hline
\hline
\multicolumn {2} {c ||} {Trans. 1 to 2} & Measured & SD3 & SM1 & SM2 & SM3 \\
\hline
\multicolumn {2} {c ||} {$t_{80}$ (s)} & $300 \pm 30$ & $263$ & $248$ & $239$ & $261$ \\
\multicolumn {2} {c ||} {$t_{50}$ (s)} & $1290 \pm 130$ & $1397$ & $1327$ & $1305$ & $1372$ \\
\multicolumn {2} {c ||} {$t_{20}$ (s)} & $3740 \pm 370$ & $4129$ & $4601$ & $4686$ & $4594$ \\
\hline
\hline
\multicolumn {2} {c ||} {Trans. 2 to 4} & Measured & SD3 & SM1 & SM2 & SM3 \\
\hline
\multicolumn {2} {c ||} {$t_{80}$ (s)} & $92 \pm 9$ & $104$ & $87$ & $81$ & $91$ \\
\multicolumn {2} {c ||} {$t_{50}$ (s)} & $250 \pm 25$ & $264$ & $273$ & $269$ & $273$ \\
\multicolumn {2} {c ||} {$t_{20}$ (s)} & $510 \pm 50$ & $410$ & $481$ & $528$ & $460$ \\
\hline
\hline
\multicolumn {2} {c ||} {Total $\chi^2$} & & $19.6$ & $25.0$ & $30.3$ & $24.1$ \\
\end{tabular}
\end{ruledtabular}
\end{table}

In the experiments, the variables $G$, $\rho_0$, and $w$ are controlled or directly determined, so the 12 independent parameters in the simplified analytical solution reduce to 9 unknown quantities ($D \rho_\mathrm{eq}^0 M_0$, $D \rho_\mathrm{eq}^0 \ell^3$, $f_\alpha^0$, and the six $D / \kappa_x^j$) to be determined or constrained by the measurements.
This is a challenge because there are only 10 measured quantities (four steady-state $\alpha$ terrace fractions $f_\alpha^\mathrm{ss}$ at different growth rates $G$, and six characteristic times for transitions in $G$.)
As described in Appendix B, in some limits the number of effective parameters is smaller, since only certain combinations of $D / \kappa_x^j$ enter the solutions.

To calculate BCF model results to compare with the experimental conditions, we assume that the only parameter affected by the Ga precursor supply rate is the deposition flux $F$, and that the only parameter affected by the presence of H$_2$ in the carrier gas is the adatom lifetime $\tau$, and that these enter only through the net growth rates $G$ determined in the experiments, given in Table \ref{tab:tableM} for each condition.
The assumption that the kinetic parameters are the same for all growth conditions is reasonable since the experiments found that the surface reconstruction did not vary over the range of conditions studied \cite{2020_Ju_NC}.
We use the experimental values $\rho_0 = 2 a^{-2} / \sqrt{3} = 1.13 \times 10^{19}$ m$^{-2}$ and $w = c / \sin(0.52^\circ) = 5.73 \times 10^{-8}$ m, where $a = 3.20 \times 10^{-10}$ m and $c = 5.20 \times 10^{-10}$ m are the lattice parameters of GaN at the growth temperature of 1073~K \cite{2000_Reeber_JMaterRes15_40}.

To explore the full range of BCF model parameters and the physics underlying them, we first searched for the best fits using the expressions obtained in Appendix B for each of the three limiting cases (diffusion-limited, attachment-limited, and mixed kinetics).
The best fit was determined by minimizing the goodness-of-fit parameter $\chi^2 \equiv \sum [(y_i - y_i^\mathrm{calc})/\sigma_i]^2$, where the $y_i$ and $\sigma_i$ are the ten measured quantities and their uncertainties.
(For this purpose the logarithms of the characteristic times were used as $y_i$.)
These initial fits are described in Appendix C.
While several of the fits give reasonable results, the parameter values obtained are not always self-consistent with the limiting cases used.

We have therefore fit the experimental data using the more general expressions from the simplified analytical solution, Eqs.~(\ref{eq:dfdta}-\ref{eq:fSSG}).
Six fits were carried out, labelled SD1, SD2, SD3 and SM1, SM2, SM3.
The starting points for fits SD1-SD3 were parameter sets close to the diffusion-limited fits D1-D3 in Appendix C, while the starting points for SM1-SM3 were parameter sets close to the mixed kinetics fits M1-M3, respectively.
From these starting points, 8 or 9 parameters were allowed to vary to find the local minimum of $\chi^2$.
As in Appendix C, we considered three functional forms for $M$: fixed $M_0 = 0$ with varying $f_\alpha^0$; fixed $f_\alpha^0 = 0.5$ with varying $M_0$; and varying both $M_0$ and $f_\alpha^0$.
The number in the fit label (1, 2, or 3) corresponds to the form used for $M$.

\begin{figure}
\includegraphics[width=0.85\linewidth]{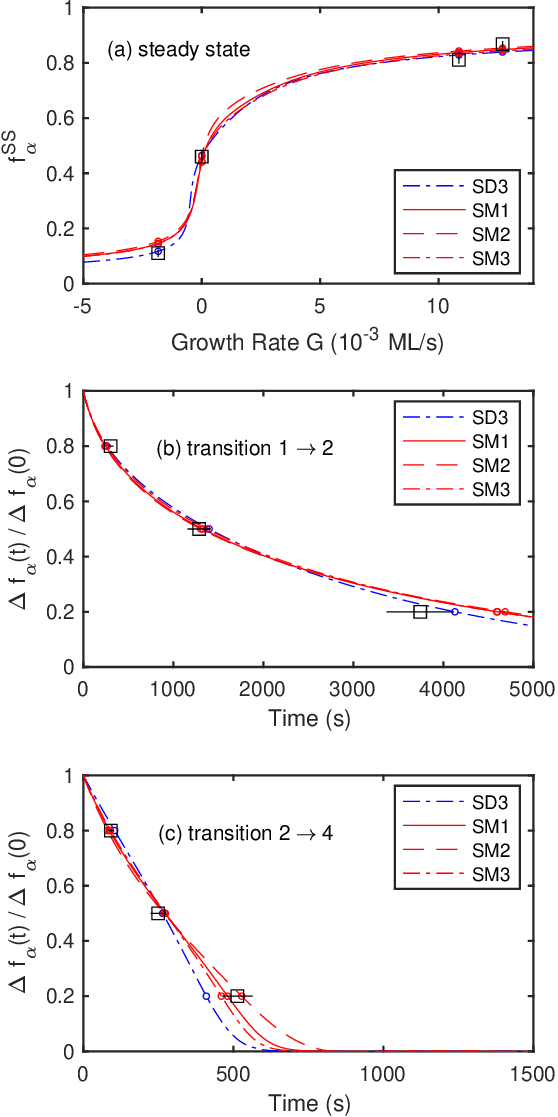}
\caption{Four best fits of the simplified analytical model from Table \ref{tab:tableY}. (a) Calculated $f_\alpha^\mathrm{ss}(G)$ curves compared with experimental $f_\alpha^\mathrm{ss}$ values (black squares). (b,c) Calculated normalized deviations $\Delta f_\alpha(t)/\Delta f_\alpha(0)$ compared with experimental characteristic times at 80\%, 50\%, and 20\% (black squares). Points on curves show calculated values compared with experimental values from Table \ref{tab:tableM} to obtain $\chi^2$. \label{fig:fss_exp}}
\end{figure}

The results are summarized in Table~\ref{tab:tableY}.
In each case, the values of some of the kinetic lengths $D/\kappa_x^j$ could be varied with no significant effect, as long as they were sufficiently smaller or larger than the terrace width $w$.
All of the fits produce $G^\mathrm{ss}(f_\alpha)$ that increase monotonically.
Figure~\ref{fig:fss_exp} compares the calculated $f_\alpha^\mathrm{ss}(G)$ curves for the four fits with the lowest $\chi^2$ to the measured points, as well as the calculated dynamics of the normalized deviations $\Delta f_\alpha(t) / \Delta f_\alpha(0)$ to the measured characteristic times.

\begin{table} 
\caption{ \label{tab:tableY} Best-fit parameter values from the simplified analytical BCF model corresponding to the diffusion-limited and mixed-kinetics cases. Also given are characteristic values $f_\alpha^*$, $G^*$, $t^*$, and $(w/\ell)^3 M_0$ for each fit. For the kinetic lengths $D/\kappa_x^j$, (small) and (large) mean much smaller or much larger than the terrace width $w = 5.73 \times 10^{-8}$~m.}
\begin{tabular}{ c || c | c | c }
\hline
\hline
\multicolumn {4} {c} {Simplified analytical near diffusion-limited} \\
\hline
Fit type: & SD1 & SD2 & SD3 \\
 & Fix $M_0 = 0$ & Fix $f_\alpha^0 = 0.5$ &  Vary Both \\
 & Vary $f_\alpha^0$ & Vary $M_0$ & $M_0$ and $f_\alpha^0$ \\
\hline
$D/\kappa_+^A$ (m) & $4.04 \times 10^{-9}$ & (small) & (large) \\
$D/\kappa_-^A$ (m) & (small) & (small) & $7.29 \times 10^{-9}$ \\
$D/\kappa_0^A$ (m) & (large) & $1.85 \times 10^{-9}$ & (large) \\
$D/\kappa_+^B$ (m) & $4.57 \times 10^{-9}$ & $2.26 \times 10^{-9}$ & $4.57 \times 10^{-8}$ \\
$D/\kappa_-^B$ (m) & $4.04 \times 10^{-9}$ & $2.44 \times 10^{-9}$ & $5.58 \times 10^{-8}$ \\
 $D/\kappa_0^B$ (m) & (small) & (small) & (small) \\
$D \rho_\mathrm{eq}^0 \ell^3$ (m$^3$ s$^{-1}$)& $7.96 \times 10^{-24}$ & $4.88 \times 10^{-24}$ & $4.44 \times 10^{-23}$   \\
$f_\alpha^0$ & $0.390$ & $0.500$ & $0.278$ \\
$D \rho_\mathrm{eq}^0 M_0$ (s$^{-1}$)& $0$ & $3.13 \times 10^{-2}$ & $-5.27 \times 10^{-1}$   \\
\hline
$f_\alpha^*$ & $0.390$ & $0.410$ & $0.464$ \\
$G^*$ (ML/s) & $0.0026$ & $0.0022$ & $0.0062$ \\
$t^*$ (s) & $4970$ & $5730$ & $2680$ \\
$(w/\ell)^3 M_0$ & $0$ & $1.21$ & $-2.23$ \\
\hline
$\chi^2$ & $59.5$ & $108.7$ & $19.6$ \\
\hline
\hline
\multicolumn {4} {c} {Simplified analytical for mixed kinetics} \\
\hline
Fit type: & SM1 & SM2 & SM3 \\
 & Fix $M_0 = 0$ & Fix $f_\alpha^0 = 0.5$ &  Vary Both \\
 & Vary $f_\alpha^0$ & Vary $M_0$ & $M_0$ and $f_\alpha^0$ \\
\hline
$D/\kappa_+^A$ (m) & (small) & (small) & (small) \\
$D/\kappa_-^A$ (m) & (large) & (large) & (large) \\
$D/\kappa_0^A$ (m) & (large) & (large) & (large) \\
$D/\kappa_+^B$ (m) & $1.82 \times 10^{-8}$ & $1.90 \times 10^{-8}$ & $1.80 \times 10^{-8}$ \\
$D/\kappa_-^B$ (m) & (large) & (large) & (large) \\
$D/\kappa_0^B$ (m) & $8.98 \times 10^{-9}$ & $1.19 \times 10^{-8}$ & $7.66 \times 10^{-9}$ \\
$D \rho_\mathrm{eq}^0 \ell^3$ (m$^3$ s$^{-1}$)& $3.58 \times 10^{-23}$ & $3.25 \times 10^{-23}$ & $3.87 \times 10^{-23}$   \\
$f_\alpha^0$ & $0.441$ & $0.500$ & $0.416$ \\
$D \rho_\mathrm{eq}^0 M_0$ (s$^{-1}$)& $0$ & $8.16 \times 10^{-2}$ & $-6.36 \times 10^{-2}$   \\
\hline
$f_\alpha^*$ & $0.441$ & $0.461$ & $0.442$ \\
$G^*$ (ML/s) & $0.0028$ & $0.0024$ & $0.0032$ \\
$t^*$ (s) & $4150$ & $4540$ & $3840$ \\
$(w/\ell)^3 M_0$ & $0$ & $0.47$ & $-0.31$ \\
\hline
$\chi^2$ & $25.0$ & $30.3$ & $24.1$ \\
\hline
\hline
\end{tabular}
\end{table}

The mixed-kinetics fits SM1-SM3 are almost identical to M1-M3 in Appendix C. All three give rather similar parameter sets and have low $\chi^2$.
The near-diffusion-limited fits SD1-SD3 give three significantly different parameter sets, in particular for the kinetic lengths of the $A$ steps.
Only fit SD3 has a low $\chi^2$, and it is the lowest of all six fits.
Table~\ref{tab:tableM} compares the experimental and calculated values for the four best fits.
The improvement of SD3 over SM1-SM3 is primarily in the fit to $f_\alpha^{ss}$ at negative $G$.

Different sets of the kinetic lengths take on limiting values (either much larger or much smaller than $w$) for fits SD1-SD3 compared with fits SM1-SM3.
All four of the fits having the best $\chi^2$ values, i.e. SD3 and SM1-SM3, give similar values of the combined parameter $D \rho_\mathrm{eq}^0 \ell^3$. 
The combined parameter $D \rho_\mathrm{eq}^0 M_0$ ranges between $-0.527$ and $+0.082$~s$^{-1}$.
To interpret these combined parameter values, we can use estimates extracted from the literature \cite{2020_Ju_NC} of $\rho_\mathrm{eq}^0 = 3.44 \times 10^{12}$~m$^{-2}$ and $D = 1.35 \times 10^{-8}$~m$^2$~s$^{-1}$ for GaN $(0 0 0 1)$ at $T = 1073$~K under similar OMVPE conditions.
For example, the parameters obtained from the SM3 fit then imply kinetic coefficients of $\kappa_+^B = 0.75$~m~s$^{-1}$ and $\kappa_0^B = 1.76$~m~s$^{-1}$, a step repulsion length of $\ell = 9.4 \times 10^{-10}$~m, and a step adatom affinity asymmetry of $M_0 = -1.3 \times 10^{-6}$. 
The example calculations shown in Figs.~\ref{fig:dfdt1}-\ref{fig:fvst1} and Table~\ref{tab:table2} correspond to the parameter values estimated in this way for the SM1 fit.

\section{Discussion and Conclusions}

The above analysis shows how the kinetic coefficients for adatom attachment and transmission at the $A$ and $B$ steps determine both the full-steady-state values $f_\alpha^\mathrm{ss}$ and the quasi-steady-state dynamics $f_\alpha(t)$ of the fraction of $\alpha$ terraces on the surface.
The exact solution can be expressed using the matrix formulas Eqs.~(\ref{eq:MC}-\ref{eq:MB}).
We obtain a simplified analytical solution for the limit in which the diffusion length of adatoms within their lifetime is much larger than the terrace width, $\sqrt{D \tau} >> w$, and the deviations of the adatom densities from their equilibrium values are small, $\rho_i / \rho_\mathrm{eq}^0 << 1$.
In this limit the evaporation flux is uniform, the net growth rate is simply proportional to the difference between the deposition and evaporation fluxes $G = (F - \rho_\mathrm{eq}^0 / \tau) / \rho_0$, and the deposition flux $F$ and adatom lifetime $\tau$ enter only in this combination.
We obtain explicit analytical expressions Eqs.~(\ref{eq:dfdta}-\ref{eq:fSSG}) for both the full-steady-state terrace fraction as a function of growth rate, $f_\alpha^\mathrm{ss}(G)$, and the dynamics, $f_\alpha(t)$, in terms of the kinetic coefficients.
The nature of $f_\alpha^\mathrm{ss}(G)$ reflects the differences in the kinetic coefficients of the $A$ and $B$ steps, and agrees qualitatively with expectations from previous work \cite{zhao2015refined,2011_Zaluska-Kotur_JAP109_023515, 2010_Zaluska-Kotur_JNoncrystSolids356_1935,2007_Sato_EurPhysJB59_311,2006_Xie_PRB_085314,2005_Frisch_PRL94_226102}.
For example, $f_\alpha^\mathrm{ss}(G)$ generally has a positive slope when the $\kappa_x^A$ are larger than the $\kappa_x^B$.
However, because there are three independent coefficients for each step type, giving six kinetic lengths $D / \kappa_x^j$ whose values relative to the terrace width affect behavior, a variety of specific cases can arise.
Diffusion-limited, attachment-limited, and mixed kinetics cases are considered in Appendix B.
For situations outside the region of validity of the simplified analytical solution, described in Supplemental Material \cite{BCF_supplemental}, the exact matrix solution can be used.

We include the effects of step transparency, which can be considered to be an artifact that arises in a 1D model to account for perpendicular transport of adatoms along the steps.
The discussion in Appendix A shows that the alternative coefficients $\kapt_x^j$ introduced in Eqs.~(\ref{eq:bca1}-\ref{eq:bca4}) are equal to the elementary attachment coefficients in Eqs.~(\ref{eq:bc1b}-\ref{eq:bc4b}) that account for the density of adatoms attached to steps.
This analysis also provides a parameter that quantitatively characterizes the transparency of the type $j$ step,
\begin{equation}
    \Theta_j = \kappa_+^j \kappa_-^j / \langle \kappa^2 \rangle .
\end{equation}
In the limits $\Theta_j \rightarrow 0$ or $\Theta_j \rightarrow 1$, the step is highly transparent or non-transparent, respectively. 

The analytical expressions obtained here can be used to fit experimental observations of $f_\alpha^\mathrm{ss}(G)$ and $f_\alpha(t)$, to elucidate the rate-limiting physical processes that underlie step-flow growth and evaporation on surfaces with alternating step types.
We present a set of fits to recent data for steady-state and dynamic values of the terrace fraction $f_\alpha$ during OMVPE growth of GaN $(0 0 0 1)$ \cite{2020_Ju_NC} obtained from analysis of \textit{in situ} surface X-ray scattering \cite{2021_Ju_PRB_CTR}.
Note that this X-ray scattering analysis assumes that the $\alpha$ and $\beta$ terraces form a sequence with a period of $m$ unit cells, with $\alpha$ terraces of $n$ unit cell width and $\beta$ terraces of $m-n$ width, so that the $\alpha$ terrace fraction is $f_\alpha = n/m$, where $n$ and $m$ are integers.
Fig.~\ref{fig:A_B} shows an example for $n = 3$, $m = 5$.
However, the apparent locations of the $A$ and $B$ steps (shown as vertical lines in Fig.~\ref{fig:A_B}) are offset from the unit cell boundaries at integer $y/b$, and the offset differs by about 1/4 unit cell for the $A$ and $B$ steps.
Using these apparent step locations, which might be more appropriate for modeling the step repulsion effects, would give a terrace fraction of $f_\alpha = (n+0.25)/m$.
We neglect this difference since the experimental terraces are many unit cells in width, $m \approx 103$.

We only consider periodically spaced steps in this paper and do not evaluate the multi-step bunching instability \cite{2020_Guin_PRL124_036101,2003_Pierre-Louis_SurfSci529_114,2017_Bellmann_JCrystGrowth478_187,2007_Dufay_PRB75_241304,2016_Li_ApplSurfSci371_242,2000_Pimpinelli_SurfSci445_L23,sato1995morphological} (apart from $A$-$B$ step pairing). 
The motivating experiments \cite{2020_Ju_NC} show that multi-step bunching does not occur under the conditions studied.

The fits of the simplified analytical solution to the experimental data are summarized in Table~\ref{tab:tableY}.
Four fits were found that give similar locally minimized values of $\chi^2$, shown in Fig.~\ref{fig:fss_exp} and Table \ref{tab:tableM}.
The best fit obtained with $M_0$ fixed at zero is the SM1 fit, indicating mixed kinetics (not completely diffusion or attachment limited) with some kinetic lengths $D/\kappa_x^j$ larger and some smaller than the terrace width.
When $M_0$ is allowed to vary, a slightly better fit is obtained with a rather different set of parameters (SD3).
There are two significantly different sets of parameter values, fit SD3 and the similar results of fits SM1-3, that reproduce the results of the experiments.
A more extensive range of experimental data, such as values of $f_\alpha^{ss}$ at additional growth rates, would be needed to better differentiate between these cases.
Interestingly, the four best fits all give the same results for the transparencies of the steps.
In all cases, $\Theta_B$ is small, indicating that the $B$ step is highly transparent, while $\Theta_A$ has an indeterminate value, so that the transparency of the $A$ step is not determined.
Likewise, the similar values of $D \rho_\mathrm{eq}^0 \ell^3$ obtained from the four fits give a consistent value for the step repulsion length of $\ell = 9.4 \pm 0.5 \times 10^{-10}$~m using estimated values of $D$ and $\rho_\mathrm{eq}^0$.

Our analysis of the experimental results assumes that the only difference between the four conditions studied is the net growth rate $G$, and that other parameters are the same.
In particular we assume that the presence of H$_2$ in the OMVPE carrier gas only affects the adatom lifetime $\tau$, to explain its observed effect on $G$.
In principle, the presence of H$_2$ could also effect kinetic parameters such as $D$ and the $\kappa_x^j$, even though the same surface reconstruction is observed under all conditions.
To address this question, further experiments are needed with more than two deposition fluxes $F$ at each H$_2$ condition (e.g. conditions giving the same $G$ with different combinations of $F$ and $\tau$) to better determine whether all $f_\alpha^\mathrm{ss}$ values collapse onto a single curve when plotted versus $G$.

Our BCF treatment introduces a new parameter $M_0$ to quantify the difference in the terrace adatom densities in equilibrium with isolated $A$ or $B$ steps.
From arguments based on adatom binding energies, discussed in Appendix A, we expect that the value of $M_0$ is close to zero for HCP-type systems.
However, when the terrace width $w$ is relatively large, the behavior of the equilibrium terrace fraction at zero growth rate $f_\alpha^*$ is extremely sensitive to even slight deviations of $M_0$ from zero.
This is because of the large multiplier $(w/\ell)^3$ in the relation between $M_0$ and $f_\alpha^*$, shown in Fig.~\ref{fig:fsM}.
While the magnitudes of $(w/\ell)^3 M_0$ given by the fits in Table~\ref{tab:tableY} are of the order of unity, the values of $M_0$ are very small, e.g. $M_0 = -1.3 \times 10^{-6}$ from the SM3 fit.
Using this value, the analysis in Appendix A implies that the sum of the step and kink attachment energies for adatoms $E_\mathrm{step}^j + E_\mathrm{kink}^j$ are the same for $A$ and $B$ steps to within $2 kT M_0 \approx 1 \times 10^{-7}$~eV.
Thus the observation of $f_\alpha^\mathrm{ss}$ close to $0.5$ when $G = 0$ provides a very sensitive test of the equality of $E_\mathrm{step}^j + E_\mathrm{kink}^j$ for $A$ and $B$ steps.
Nonetheless even such a tiny difference has an observable effect.

The BCF analysis developed here predicts that for some combinations of material parameters,
$G^\mathrm{ss}(f_\alpha)$ can be non-monotonic, giving full-steady-state terrace fractions $f_\alpha^\mathrm{ss}(G)$ with multiple stable (and unstable) solutions in some region of $G$.
For example, in both the attachment-limited and mixed kinetics cases, $K^\mathrm{ss}(f_\alpha)$ can change sign as a function of $f_\alpha$, which typically leads to this situation.
Likewise our analysis predicts that the dynamics of $f_\alpha(t)$ after a change in condition can have a significantly non-exponential behavior, as shown in Fig.~\ref{fig:fvst1}.
While neither effect is strongly apparent in the experimental results discussed here, their appearance in future experiments could provide insight into the atomic-scale mechanisms.

\begin{acknowledgments}
  Work supported by the U.S Department of Energy (DOE), Office of Science, Office of Basic Energy Sciences, Materials Science and Engineering Division.
  Measured values shown are from experiments at beamline 12ID-D of the Advanced Photon Source, a DOE Office of Science user facility operated by Argonne National Laboratory.
\end{acknowledgments}

\appendix

\section{Step transparency and adatom diffusion along steps}

The last term in Eqs.~(\ref{eq:bc1}-\ref{eq:bc4}) accounts for step transparency \cite{2003_Pierre-Louis_PRE68_021604,2003_Pierre-Louis_SurfSci529_114,2005_Krug_ISNM149_69,2007_Ranguelov_PRB75_245419}, a phenomenon in which adatoms cross the step to exchange between neighboring terraces without attachment at a kink site on the step.
This process involves temporary adatom attachment from a terrace onto a step and some diffusion along the step, but with adatom detachment onto the opposite terrace before a kink is encountered.
Thus we can better understand step transparency by considering the density $\rho_L^j(x)$ of adatoms attached to steps and how it varies in the $x$ direction along a step of type $j$. 
Here we develop a simple model of line diffusion of adatoms along a step between kinks that couples to the model presented above for surface diffusion on terraces, to relate the kinetic coefficients $\kappa_+^j$, $\kappa_-^j$, and $\kappa_0^j$ to the line diffusivity, kink attachment coefficients, and kink density.
Models for diffusion of adatoms attached to steps have been presented previously \cite{1999_Caflisch_PRE59_6879,2004_Filimonov_SurfSci553_133,2005_Balykov_PRE72_022601}, including discrete two-dimensional models with kinks on steps \cite{ackerman2011boundary,zhao2015refined,2016_Zhao_PRB93_165411}. 
Here we couple orthogonal one-dimensional step and terrace diffusion models via the boundary conditions at steps for adatom diffusion on the terraces, Eqs.~(\ref{eq:bc1}-\ref{eq:bc4}).
Since our model of terrace diffusion allows variation only in the $y$ direction normal to the steps, it couples to the model for adatoms attached to steps through the average value $\langle \rho_L^j \rangle$ on the step.
These considerations give physical interpretations to the modified coefficients $\kapt_x^j$ and $\rhot_\mathrm{eq}^j$ in the alternative terrace boundary conditions Eqs.~(\ref{eq:bca1}-\ref{eq:bca4}).
They also can be used to obtain expressions for $\rho_\mathrm{eq}^0$ and $M_0$ in terms of the adatom attachment energies at steps and kinks.
These results apply to the standard BCF theory for surfaces with only one type of step, as well as the extension developed here for surfaces with alternating step types.

\subsection{Terrace boundary conditions with step adatom densities}

We start by re-writing the terrace boundary conditions in a form that explicitly accounts for the average density of adatoms attached to steps,
\begin{align}
    J_\alpha^+ &= - D \nabla_y \rho_\alpha^+
    = +k_-^{A} \, \rho_\alpha^+ - k_-^{Ad} \, \langle \rho_L^A \rangle \, \rho_{0y}, \label{eq:bc1a} \\
    J_\alpha^- &= - D \nabla_y \rho_\alpha^-
    = -k_+^{B} \, \rho_\alpha^- + k_+^{Bd} \, \langle \rho_L^B \rangle \, \rho_{0y}, \label{eq:bc2a} \\
    J_\beta^+ &= - D \nabla_y \rho_\beta^+
    = +k_-^{B} \, \rho_\beta^+ - k_-^{Bd} \, \langle \rho_L^B \rangle \, \rho_{0y}, \label{eq:bc3a} \\
    J_\beta^- &= - D \nabla_y \rho_\beta^-
    = -k_+^{A} \, \rho_\beta^- + k_+^{Ad} \, \langle \rho_L^A \rangle \, \rho_{0y}, \label{eq:bc4a}
\end{align}
where the coefficients $k_-^{j}$ and $k_-^{jd}$ are elementary kinetic coefficients for adatom attachment and detachment, respectively, to a step of type $j$ from a terrace on the uphill side, $k_+^{j}$ and $k_+^{jd}$ are elementary kinetic coefficients for adatom attachment and detachment, respectively, to a step from a terrace on the downhill side, $\langle \rho_L^j \rangle$ is the mean linear density of adatoms attached to the step of type $j$, and $\rho_{0y}$ is the linear density of sites perpendicular to the step (in the $y$ direction).
As before, the $+$ or $-$ superscripts on $J_i$, $\rho_i$, and $\nabla_y \rho_i$ indicate evaluation at the terrace boundaries $y = + w_i/2$ or $y = - w_i/2$, respectively.
The use of mean adatom densities on each step $\langle \rho_L^j \rangle$ is justified under the assumption that the kink spacing is much smaller than the terrace width, so that non-uniformity along the step can be neglected and the terrace transport remains a nearly one-dimensional problem.
The linear densities $\rho_L^j$ and $\rho_{0y}$ have dimensions of (length)$^{-1}$, while the areal densities $\rho_i$ have dimensions of (length)$^{-2}$.
Regarding the sign notations in Eqs.~(\ref{eq:bc1}-\ref{eq:bc4}) and Eqs.~(\ref{eq:bc1a}-\ref{eq:bc4a}), note that while we use the same notation as in most of the literature \cite{2000_Gillet_EurPhysJB18_519,2003_Pierre-Louis_PRE68_021604,2003_Pierre-Louis_SurfSci529_114,2004_Pierre-Louis_PRL93_165901,2005_Krug_ISNM149_69,2010_Patrone_PRE82_061601,2017_Bellmann_JCrystGrowth478_187,2020_Guin_PRL124_036101} for the subscripts on the kinetic coefficients (i.e. $\kappa_+^j$ for attachment from below and $\kappa_-^j$ for attachment from above), we use the opposite notation as in much of the literature \cite{2000_Gillet_EurPhysJB18_519,2003_Pierre-Louis_PRE68_021604,2003_Pierre-Louis_SurfSci529_114,2004_Pierre-Louis_PRL93_165901,2005_Krug_ISNM149_69,2010_Patrone_PRE82_061601,2020_Guin_PRL124_036101} for the superscripts on $J_i$, $\rho_i$, and $\nabla_y \rho_i$ (since here evaluation at $y = + w_i/2$ is the boundary above a step, and $y = - w_i/2$ is the boundary below a step).

To obey detailed balance, all fluxes must be zero at equilibrium.
At equilibrium we have
\begin{align}
\langle \rho_L^j \rangle &= \rho_{L\mathrm{eq}}^j, \label{eq:rhoLeq}\\
\rho_\alpha^+ &= \rho_\beta^- = \rho_\mathrm{eq}^A, \label{eq:rhoeqA}
\\
\rho_\alpha^- &= \rho_\beta^+ = \rho_\mathrm{eq}^B, \label{eq:rhoeqB}
\end{align}
where $\rho_{L\mathrm{eq}}^j$ is the equilibrium adatom linear density on the type $j$ step. 
Detailed balance then relates the detachment and attachment coefficients by
\begin{equation}
    \frac{k_-^{jd}}{k_-^{j}} = \frac{k_+^{jd}}{k_+^{j}} = \frac{\rho_\mathrm{eq}^j}{\rho_{L\mathrm{eq}}^j \rho_{0y}}. \label{eq:dbt}
\end{equation}

Using this to eliminate the detachment coefficients, the boundary conditions become
\begin{align}
    J_\alpha^+ &= - D \nabla_y \rho_\alpha^+
    = +k_-^{A} \, (\rho_\alpha^+ - \rho_\mathrm{eq}^A \, \langle \rho_L^A \rangle / \rho_{L\mathrm{eq}}^A), \label{eq:bc1b} \\
    J_\alpha^- &= - D \nabla_y \rho_\alpha^-
    = -k_+^{B} \, (\rho_\alpha^- - \rho_\mathrm{eq}^B \, \langle \rho_L^B \rangle / \rho_{L\mathrm{eq}}^B), \label{eq:bc2b} \\
    J_\beta^+ &= - D \nabla_y \rho_\beta^+
    = +k_-^{B} \, (\rho_\beta^+ - \rho_\mathrm{eq}^B \, \langle \rho_L^B \rangle / \rho_{L\mathrm{eq}}^B), \label{eq:bc3b} \\
    J_\beta^- &= - D \nabla_y \rho_\beta^-
    = -k_+^{A} \, (\rho_\beta^- - \rho_\mathrm{eq}^A \, \langle \rho_L^A \rangle / \rho_{L\mathrm{eq}}^A). \label{eq:bc4b}
\end{align}
The six coefficients in this form of the boundary conditions, $k_+^j$, $k_-^j$, and $\rho_\mathrm{eq}^j \, \langle \rho_L^j \rangle / \rho_{L\mathrm{eq}}^j$, for $j = A$ and $B$, can be related to the six kinetic coefficients in Eqs.~(\ref{eq:bc1}-\ref{eq:bc4}), $\kappa_+^j$, $\kappa_-^j$, and $\kappa_0^j$,
as well as the six kinetic coefficients in the alternative boundary conditions, Eqs.~(\ref{eq:bca1}-\ref{eq:bca4}), $\kapt_+^j$, $\kapt_-^j$, and $\rhot_\mathrm{eq}^j$. 
In the latter case the relations are especially simple,
\begin{align}
    \kapt_+^j &= k_+^j, \label{eq:kta1} \\
    \kapt_-^j &= k_-^j, \label{eq:kta2} \\
    \rhot_\mathrm{eq}^j &= \rho_\mathrm{eq}^j \, \langle \rho_L^j \rangle / \rho_{L\mathrm{eq}}^j. \label{eq:kta3} 
\end{align}
This gives a physical meaning to the alternative coefficients $\kapt_x^j$ introduced for mathematical reasons in Eqs.~(\ref{eq:bca1}-\ref{eq:bca4}). 
The $\kapt_x^j$ are equal to the elementary attachment coefficients in Eqs.~(\ref{eq:bc1a}-\ref{eq:bc4a}) and Eqs.~(\ref{eq:bc1b}-\ref{eq:bc4b}) that account for the density of adatoms attached to steps.
The density $\rhot_\mathrm{eq}^j$ is the adatom density on the terraces in equilibrium with steps having average adatom densities $\langle \rho_L^j \rangle$ that can differ from the value in equilibrium with kinks, $\rho_{L\mathrm{eq}}^j$, and depend on growth rate.

\subsection{Calculation of adatom density on steps}

\begin{figure}
\includegraphics[width=0.9\linewidth]{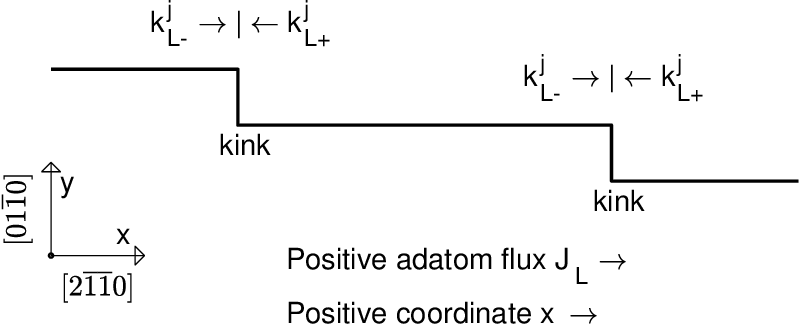}
\caption{Schematic of kinks on a step of type $j$, showing kinetic coefficients for adatom attachment at kinks. Note all kinks are identical on a given type step in this model. \label{fig:BCFL}}
\end{figure}

To calculate the mean adatom linear density ratios $\langle \rho_L^j \rangle / \rho_{L\mathrm{eq}}^j$, we can write a one-dimensional model in the $x$ direction (along the steps) analogous to the above one-dimensional model in the $y$ direction (perpendicular to the steps), where the kinks that bound straight step segments play the role of the steps that bound the terraces. 
As shown in Fig.~\ref{fig:BCFL}, we assume that that the kinks all have the same sign and are uniformly spaced by the amount needed to satisfy the geometrical requirement imposed by the overall step direction.
This can be analyzed in terms of the probabilities $n_+$ and $n_-$ for positive or negative kinks to occur at each lattice site on the step.
For a close-packed surface, the geometrical requirement gives
\begin{equation}
n_+ - n_- = 2 / (\sqrt{3}/ \tan \phi + 1), 
\end{equation}
where $\phi$ is the angle of the step with respect to the atomic rows in the $[\overline{2} 1 1 0]$ type directions.
The geometrically required average kink spacing is $a / (n_+ - n_-)$, where $a$ is the lattice parameter.

The density of kinks on a step can be larger due to additional thermally generated kink pairs \cite{1951_Burton_PhilTransRS243_29}.
The kink probabilities must satisfy
\begin{equation}
    n_+ n_- = \exp(-E_\mathrm{pair} / kT),
\end{equation}
where $E_\mathrm{pair}$ is the energy cost to generate a kink pair, and we assume the kink probabilities are much smaller than unity.
For simplicity we assume $E_\mathrm{pair} / kT >> 1$ and neglect kink pairs generated thermally or by nucleation from adatoms meeting in the step, so that all kinks have the same sign.
Thus we model ``kink flow'' growth on a step with a finite angle $\phi$.

We can write continuity equations for adatom transport on the straight step segments between the kinks,
\begin{equation}
    \frac{\partial \rho_L^j}{\partial t} = D_L^j\, \nabla_x^2 \rho_L^j + \rho_0 v_j,
    \label{eq:contL}
\end{equation}
where $D_L^j$ is the line diffusivity for step type $j$, and the last term is the adatom source/sink term from the two neighboring terraces.

The line flux boundary conditions at the kinks can be written as
\begin{align}
    J_L^{j+} &= - D_L^j \, \nabla_x \rho_L^{j+}
    = +k_{L-}^{j} \, \rho_L^{j+} - k_{L-}^{jd} \, \rho_{0x}, \label{eq:bcL1a} \\
    J_L^{j-} &= - D_L^j \, \nabla_x \rho_L^{j-}
    = -k_{L+}^{j} \, \rho_L^{j-} + k_{L+}^{jd} \, \rho_{0x}, \label{eq:bcL2a}
\end{align}
where $J_L^j$ is the adatom line flux along a step of type $j = A$ or $B$, the coefficients $k_{L-}^{j}$ and $k_{L-}^{jd}$ are elementary kinetic coefficients for adatom attachment and detachment, respectively, to a kink from the ``uphill'' side, $k_{L+}^{j}$ and $k_{L+}^{jd}$ are elementary kinetic coefficients for adatom attachment and detachment, respectively, to a kink from the ``downhill'' side, and $\rho_{0x}$ is the linear density of sites along the step (in the $x$ direction).
The $+$ or $-$ superscripts on $J_L^j$, $\rho_L^j$, and $\nabla_x \rho_L^j$ indicate evaluation at the terrace boundaries $x = + d_j/2$ or $x = - d_j/2$, respectively, where $d_j$ is the distance between kinks on steps of type $j$ and the spatial coordinate $x$ is taken to be zero in the center of the step segment. 

To obey detailed balance, all fluxes must be zero at equilibrium.
At equilibrium we have
\begin{equation}
\rho_L^{j+} = \rho_L^{j-} = \rho_{L\mathrm{eq}}^j. 
\label{eq:rhoeqL}
\end{equation}
Detailed balance then relates the detachment and attachment coefficients by
\begin{equation}
    \frac{k_{L-}^{jd}}{k_{L-}^{j}} = \frac{k_{L+}^{jd}}{k_{L+}^{j}} = \frac{\rho_{L\mathrm{eq}}^j}{\rho_{0x}}.
\end{equation}
Using this to eliminate the detachment coefficients, the boundary conditions become
\begin{align}
    J_L^{j+} &= - D_L^j \, \nabla_x \rho_L^{j+}
    = +k_{L-}^{j} \, (\rho_L^{j+} - \rho_{L\mathrm{eq}}^j), \label{eq:bcL1b} \\
    J_L^{j-} &= - D_L^j \, \nabla_x \rho_L^{j-}
    = -k_{L+}^{j} \, (\rho_L^{j-} - \rho_{L\mathrm{eq}}^j). \label{eq:bcL2b}
\end{align}
A standard positive kink Ehrlich-Schwoebel effect \cite{1999_Pierre-Louis_PRL82_3661} is given by $k_{L+}^{j} > k_{L-}^{j}$.

The kink velocity on a type $j$ step can be obtained from the adatom flux arriving from each side,
\begin{equation}
    v_\mathrm{kink}^j = (J_L^{j+} - J_L^{j-}) / \rho_{0x}.
    \label{eq:vkink}
\end{equation}
Note that we have neglected advective terms due to the velocity of the kinks, and will check the self-consistency of this assumption below.

At quasi-steady-state, the general solution for the distribution of adatoms on the step $\rho_L^j(x)$ satisfying Eq.~(\ref{eq:contL}) with $\partial \rho_L^j / \partial t = 0$ is a quadratic function
\begin{equation}
    \rho_L^j = a_j x^2 + b_j x + c_j,
\end{equation}
with derivatives
\begin{align}
    \nabla_x \rho_L^j &= 2 a_j x + b_j, \\
    \nabla_x^2 \rho_L^j &= 2 a_j.
\end{align}
By substituting these into the boundary conditions, we can solve for the coefficients to obtain
\begin{align}
    a_j &= - \frac{\rho_0 v_j}{2 D_L^j}, \\
    b_j &= \frac{\rho_0 v_j \, (k_{L+}^j - k_{L-}^j) \, d_j}{2[D_L^j (k_{L+}^j + k_{L-}^j) + k_{L+}^j \, k_{L-}^j \, d_j]}, \\
    c_j &= \rho_{L\mathrm{eq}}^j + \frac{\rho_0 v_j \, d_j \, [D_L^j + (k_{L+}^j + k_{L-}^j) d_j/4]}{D_L^j \, (k_{L+}^j + k_{L-}^j) + k_{L+}^j \, k_{L-}^j \, d_j}.
\end{align}
We can evaluate the mean adatom linear density as
\begin{align}
    \langle \rho_L^j \rangle &= \frac{1}{d_j} \int_{-d_j/2}^{d_j/2} \rho_L^j dx \nonumber \\
    &= a_j d_j^2 / 12 + c_j\nonumber \\
    &= \rho_{L\mathrm{eq}}^j + \frac{\rho_0 v_j d_j^2}{D_L^j} \times \label{eq:mean_j} \\
    & \left ( \frac{1}{12} + \frac{1 + d_j (k_{L+}^j + k_{L-}^j)/(4 D_L^j)}{d_j (k_{L+}^j + k_{L-}^j)/D_L^j + d_j^2 \, k_{L+}^j \, k_{L-}^j \, / (D_L^j)^2} \right ). \nonumber
\end{align}
Thus the deviation of $\langle \rho_L^j \rangle$ from $\rho_{L\mathrm{eq}}^j$ is proportional to the net influx of adatoms to the step, $\rho_0 v_j$.
The proportionality depends in a complex way on the line diffusivity $D_L^j$, the kink kinetic coefficients $k_{L+}^j$ and $k_{L-}^j$, and the kink spacing $d_j$.

The terrace boundary conditions Eqs.~(\ref{eq:bc1b}-\ref{eq:bc4b}) give a second relationship between $\langle \rho_L^j \rangle$ and $\rho_0 v_j$ for each step type $j = A$ or $B$,
\begin{align}
    \langle \rho_L^A \rangle &= \rho_{L\mathrm{eq}}^A \left ( \frac{k_-^A \, \rho_\alpha^+ + k_+^A \, \rho_\beta^- - \rho_0 v_A}{(k_-^A + k_+^A) \, \rho_\mathrm{eq}^A} \right ), \label{eq:mean_A}\\
    \langle \rho_L^B \rangle &= \rho_{L\mathrm{eq}}^B \left ( \frac{k_-^B \, \rho_\beta^+ + k_+^B \, \rho_\alpha^- - \rho_0 v_B}{(k_-^B + k_+^B) \, \rho_\mathrm{eq}^B} \right ), \label{eq:mean_B}
\end{align}
where we have used Eqs.~(\ref{eq:va1}-\ref{eq:vb1}) between the fluxes and the $v_j$.
By setting these relations equal to those from the step solution, Eq.~(\ref{eq:mean_j}), and eliminating $\rho_0 v_j$, we can obtain expressions for the mean adatom linear density ratios,
\begin{align}
    \frac{\langle \rho_L^A \rangle}{\rho_{L\mathrm{eq}}^A} &= \Theta_A + (1 - \Theta_A) \left ( \frac{k_-^A \, \rho_\alpha^+ + k_+^A \, \rho_\beta^-}{(k_-^A + k_+^A) \, \rho_\mathrm{eq}^A} \right ), \label{eq:mean_A2} \\
    \frac{\langle \rho_L^B \rangle}{\rho_{L\mathrm{eq}}^B} &= \Theta_B + (1 - \Theta_B) \left ( \frac{k_-^B \, \rho_\beta^+ + k_+^B \, \rho_\alpha^-}{(k_-^B + k_+^B) \, \rho_\mathrm{eq}^B} \right ), \label{eq:mean_B2}
\end{align}
where we have defined the fraction $\Theta_j$ by
\begin{align}
    &\Theta_j = \frac{R_\mathrm{incorp}}{R_\mathrm{incorp} + R_\mathrm{detach}}, \\
    &R_\mathrm{incorp} \equiv \frac{D_L^j}{d_j^2} \times \\
    & \left ( \frac{1}{12} + \frac{1 + d_j (k_{L+}^j + k_{L-}^j)/(4 D_L^j)}{d_j (k_{L+}^j + k_{L-}^j)/D_L^j + d_j^2 \, k_{L+}^j \, k_{L-}^j \, / (D_L^j)^2} \right )^{-1}, \nonumber\\
    &R_\mathrm{detach} \equiv (k_+^j + k_-^j) \, \rho_\mathrm{eq}^j \, / \, \rho_{L\mathrm{eq}}^j .
\end{align}
The fraction $\Theta_j$ varies between zero and unity, and can be treated as the probability that an adatom attached to a step incorporates at a kink that it reaches via diffusion along the step, rather than detaching from the step onto a neighboring terrace \cite{2004_Filimonov_SurfSci553_133}.
The incorporation rate per unit step adatom density $R_\mathrm{incorp}$ can be limited by diffusion to a kink, attachment at a kink, or a combination, depending upon the relative values of the kinetic lengths $D_L^j / k_{L+}^j$ and $D_L^j / k_{L-}^j$ and the kink spacing $d_j$. 
The detachment rate is the flux per unit $\langle \rho_L^j \rangle$, given by $R_\mathrm{detach} = (k_-^{jd} + k_+^{jd}) \rho_{0y}$, which can be evaluated using Eq.~(\ref{eq:dbt}).

If we substitute expressions (\ref{eq:mean_A2}) and (\ref{eq:mean_B2}) for $\langle \rho_L^j \rangle$ into the terrace boundary conditions Eqs.~(\ref{eq:bc1b}-\ref{eq:bc4b}) and then equate these to the original terrace boundary conditions Eqs.~(\ref{eq:bc1}-\ref{eq:bc4}), we obtain expressions for the kinetic coefficients in the original boundary conditions,
\begin{align}
    \kappa_+^j &= \Theta_j \, k_+^j, \label{eq:kkp} \\
    \kappa_-^j &= \Theta_j \, k_-^j, \label{eq:kkn} \\
    \kappa_0^j &= \frac{(1 - \Theta_j) \, k_+^j \, k_-^j }{k_+^j + k_-^j}, \label{eq:kk0} \\
    \langle \kappa^2 \rangle ^j &= \Theta_j \, k_+^j \, k_-^j.
\end{align}
Thus the fraction $\Theta_j$ determines the transparency of the step of type $j$; when $\Theta_j$ approaches unity, e.g. large $R_\mathrm{incorp}$, the step is non-transparent $(\kappa_0^j << \kappa_+^j, \kappa_-^j)$, while when $\Theta_j$ approaches zero, e.g. large $R_\mathrm{detach}$, the step can be highly transparent $(\kappa_0^j >> \kappa_+^j, \kappa_-^j)$.

One can invert these expressions to obtain
\begin{align}
    k_+^j &= \langle \kappa^2 \rangle ^j \, / \, \kappa_-^j, \\
    k_-^j &= \langle \kappa^2 \rangle ^j \, / \, \kappa_+^j, \\
    \Theta_j &= \kappa_+^j \, \kappa_-^j \, / \, \langle \kappa^2 \rangle ^j.
\end{align}
Comparing these to the definitions of the coefficients in the alternative boundary conditions, Eqs.~(\ref{eq:kt1}-\ref{eq:kt3}), one obtains Eqs.~(\ref{eq:kta1}-\ref{eq:kta3}) and
\begin{align}
    \rhot_\mathrm{eq}^j &= \rho_\mathrm{eq}^j + \frac{(1 - \Theta_j) \, \rho_0 v_j}{\Theta_j \, (k_+^j + k_-^j)} \\
    &= \rho_\mathrm{eq}^j \, \left ( 1 + \frac{\rho_0 v_j}{\rho_{L\mathrm{eq}}^j \, R_\mathrm{incorp}} \right ). \nonumber
\end{align}

Two previous treatments \cite{1999_Pierre-Louis_PRL82_3661,2004_Filimonov_SurfSci553_133} have evaluated diffusion of adatoms along steps to obtain expressions for the kinetic coefficients including the effects of step transparency.
Both considered only a single type of step.
The first treatment \cite{1999_Pierre-Louis_PRL82_3661} used a continuity equation for adatoms attached to steps similar to Eq.~(\ref{eq:contL}) but with an additional loss term due to a uniform kink density, $\nu (\rho_L - \rho_{L\mathrm{eq}})$, with a kinetic coefficient $\nu$.
The continuity equation and boundary conditions for terrace adatom transport were the same as Eqs.~(\ref{eq:cont}) and (\ref{eq:bc1b}-\ref{eq:bc4b}).
Expressions for the kinetic coefficients were obtained only for a uniform adatom density on the step $\rho_L = \langle \rho_L \rangle$.
As in Eq.~(\ref{eq:mean_j}), this gives a deviation of $\langle \rho_L \rangle$ from $\rho_{L\mathrm{eq}}$ proportional to the net influx of adatoms to the step, $\rho_0 v = \nu (\langle \rho_L \rangle$ - $\rho_{L\mathrm{eq}})$.
The expressions obtained for the kinetic coefficients are equivalent to Eqs.~(\ref{eq:kkp}-\ref{eq:kk0}) if we identify $\nu$ as
\begin{equation}
    \nu = \frac{\Theta \, (k_+ + k_-) \, \rho_\mathrm{eq}}{(1 - \Theta) \, \rho_{L\mathrm{eq}}}.
\end{equation}
The second treatment \cite{2004_Filimonov_SurfSci553_133} used a continuity equation for adatoms attached to steps equivalent to Eq.~(\ref{eq:contL}) with boundary conditions at kinks equivalent to Eqs.~(\ref{eq:bcL1b}-\ref{eq:bcL2b}).
It considered only the case $k_{L-} = k_{L+}$ (no kink Ehrlich-Schwoebel effect).
The expressions obtained for the kinetic coefficients are equivalent to Eqs.~(\ref{eq:kkp}-\ref{eq:kk0}) but with a different formula for $\Theta$.
In our notation their formula is
\begin{align}
    \Theta &= \frac{\tanh{Q}}{Q \left ( 1 + \frac{2 D_L}{d \, k_L} Q \tanh{Q} \right )}, \label{eq:TQ1} \\
    Q &\equiv \frac{d}{2} \left [ \frac{(k_+ + k_-) \rho_\mathrm{eq}}{D_L \, \rho_{L\mathrm{eq}}} \right ]^{1/2}.
\end{align}
In this limit, our formula for $\Theta$ can be written as
\begin{equation}
    \Theta = \frac{1}{1 + \left ( \frac{1}{3} + \frac{2 D_L}{d \, k_L} \right ) Q^2 }. \label{eq:TQ2}
\end{equation}
The two formulas (\ref{eq:TQ1}) and (\ref{eq:TQ2}) have similar behavior, with identical limits for $Q \rightarrow 0$ and $Q \rightarrow \infty$ when the $1/3$ term in Eq.~(\ref{eq:TQ2}) is negligible.
However, in the limit of large $Q$ and small $D_L/(d k_L)$ (e.g. diffusion-limited step transport), Eq.~(\ref{eq:TQ1}) reduces to $\Theta = Q^{-1} [1 + D_L Q / (d k_L)]^{-1}$, while Eq.~(\ref{eq:TQ2}) gives $\Theta = 3 Q^{-2}$.
Although these both approach zero at large $Q$, the detailed derivation provided above suggests that the latter is more accurate.

To evaluate the advective contribution to adatom transport on steps, we note that at quasi-steady-state, the divergence of the line flux is a constant,
\begin{equation}
    \nabla_x \cdot J_L^j = -D_L^j \, \nabla_x^2 \rho_L^j = \rho_0 v_j.
\end{equation}
The kink velocity of Eq.~(\ref{eq:vkink}) can be expressed as 
\begin{equation}
    v_\mathrm{kink}^j = d_j \nabla_x \cdot J_L^j / \rho_{0x} = d_j \rho_{0y} \, v_j,
\end{equation}
where we have made use of the relation $\rho_0 = \rho_{0x} \rho_{0y}$.
The kink and step velocities are related to each other and the growth rate via
\begin{align}
     \rho_0 G w &= \rho_{0x} \left ( \frac{v_\mathrm{kink}^A}{d_A} + \frac{v_\mathrm{kink}^B}{d_B} \right ) \nonumber \\
     &= \rho_{0x} \rho_{0y} (v_A + v_B).
\end{align}
The advective contributions to the line fluxes $\rho_L^j \, v_\mathrm{kink}^j = \rho_L^j \, d_j \, \rho_{0y} v_j$ are thus always a small fraction of the line flux obtained by integrating the divergence $d_j \, \nabla_x \cdot J_L^j = d_j \rho_0 v_j$ when the adatom coverage on the step is small, $\rho_L^j << \rho_{0x}$.

\subsection{Adatom binding energies}

We can relate $\rho_\mathrm{eq}^0$ and $M_0$ to the binding energies for adatoms at steps and kinks.
Such adatom binding energies at steps (but not kinks) have been calculated for GaN $(0 0 0 1)$ in OMVPE conditions \cite{2020_Akiyama_JCrystGrowth532_125410,2020_Akiyama_JJAP59_SGGK03,ohka2020effect}.
If we assume isolated steps and neglect step-step and kink-kink interactions, the equilibria between adatom densities on steps and terraces can be written as
\begin{align}
    \rho_{L\mathrm{eq}}^j &= \rho_{0x} \,  \exp(-E_\mathrm{kink}^j/kT), \\
    \rho_\mathrm{eq}^j &= \rho_{L\mathrm{eq}}^j \, \rho_{0y} \, \exp(-E_\mathrm{step}^j/kT),
\end{align}
where $E_\mathrm{step}^j$ is the binding energy of a terrace adatom to a step of type $j$, and $E_\mathrm{kink}^j$ is the binding energy of an adatom attached to a step of type $j$ to a kink.
These combine to give
\begin{equation}
    \rho_\mathrm{eq}^j = \rho_0 \exp \left ( - \frac{E_\mathrm{step}^j + E_\mathrm{kink}^j}{kT} \right ),
\end{equation}
where the sum $E_\mathrm{step}^j + E_\mathrm{kink}^j$ is the total energy for a terrace adatom to bind to a kink.
From Eqs.~(\ref{eq:rhoeqj}) and (\ref{eq:muAB}) we obtain
\begin{align}
    \rho_\mathrm{eq}^0 &= (\rho_\mathrm{eq}^A \rho_\mathrm{eq}^B)^{1/2} \\
    &= \rho_0 \exp \left ( - \frac{E_\mathrm{step}^A + E_\mathrm{kink}^A + E_\mathrm{step}^B + E_\mathrm{kink}^B}{2kT} \right ), \nonumber \\
    M_0 &= \frac{1}{2} (\log{\rho_\mathrm{eq}^A} - \log{\rho_\mathrm{eq}^B}) \\ 
    &= \frac{E_\mathrm{step}^B + E_\mathrm{kink}^B - (E_\mathrm{step}^A + E_\mathrm{kink}^A)}{2kT}. \nonumber
\end{align}

Nearest-neighbor bond-counting arguments can be used to give $E_\mathrm{step}^j + E_\mathrm{kink}^j = E_\mathrm{bulk} - E_\mathrm{ads}$, where $E_\mathrm{bulk}$ is the bulk cohesive energy of the crystal per atom, and $E_\mathrm{ads}$ is the adsorption energy of an adatom on the terrace below the step.
For HCP-type systems, where we expect $E_\mathrm{ads}$ to be the same for $\alpha$ and $\beta$ terraces, this argument gives $E_\mathrm{step}^A + E_\mathrm{kink}^A = E_\mathrm{step}^B + E_\mathrm{kink}^B$, or $M_0 = 0$.
The estimated value of $\rho_\mathrm{eq}^0 = 3.44 \times 10^{12}$~m$^{-2}$ for GaN $(0 0 0 1)$ at $T = 1073$~K in OMVPE conditions \cite{2020_Ju_NC} and the value of $\rho_0 = 1.13 \times 10^{19}$~m$^{-2}$ gives a value of $E_\mathrm{step}^A + E_\mathrm{kink}^A = E_\mathrm{step}^B + E_\mathrm{kink}^B = 1.39$~eV.

\section{Limiting cases of the simplified analytical solution}

Here we show how the expressions developed above for the simplified analytical solution reduce to simpler expression for cases in which the adatom kinetics on the terraces are limited by diffusion or by attachment/detachment at steps.
For each, we consider the sub-cases of non-transparent or highly transparent steps, and examine the factors that determine the sign of $K^\mathrm{ss}$, and thus whether $f_\alpha^\mathrm{ss}(G)$ has a positive or negative slope.
We finally consider a mixed case in which $\alpha$ and $\beta$ terraces have different limiting kinetics.

\subsection{Diffusion-limited kinetics}

In the diffusion-limited case, the first two terms are negligible in Eq.~(\ref{eq:Ra}) for $R_\alpha$ and in Eq.~(\ref{eq:Rb}) for $R_\beta$.
These expressions reduce to $R_\alpha = f_\alpha^{-1}$ and $R_\beta = (1 - f_\alpha)^{-1}$.
The coefficients $S_\alpha$ and $S_\beta$ become independent of $f_\alpha$.
The expression for $K^\mathrm{ss}$ is given by
\begin{equation}
    K^\mathrm{ss} \approx \big [ W_0^{dl} + W_1^{dl} f_\alpha (1 - f_\alpha) \big ]^{-1},
    \label{eq:Dk2}
\end{equation}
where we have introduced coefficients
\begin{align}
    W_0^{dl} &\equiv \frac{\kappa_0^B}{\kapsq^B} - \frac{\kappa_0^A}{\kapsq^A}, \\
    W_1^{dl} &\equiv \frac{\kappa_+^B}{\kapsq^B} + \frac{\kappa_-^B}{\kapsq^B} - \frac{\kappa_+^A}{\kapsq^A} - \frac{\kappa_-^A}{\kapsq^A} .
\end{align}
Since the $\kappa_x^j$ must all be positive, the values of these coefficients obey the limits $|W_0^{dl}| \leq w R_0 / D$ and $|W_1^{dl}| << w/D$, where the second relation is based on the diffusion-limited approximation. 
The expression for $K^\mathrm{dyn}$ becomes
\begin{equation}
    K^\mathrm{dyn} \approx \frac{D}{w [f_\alpha (1 - f_\alpha) + R_0]}. \label{eq:dfdta2}
\end{equation}

For the sub-case of non-transparent steps, with $\kappa_0^A = \kappa_0^B = 0$, we have $\kapsq^j = \kappa_+^j \kappa_-^j$ for both steps $j = A$ and $B$.
The expression for $K^\mathrm{ss}$ becomes
\begin{equation}
    K^\mathrm{ss} \approx
    \left [ f_\alpha (1 - f_\alpha) \left ( \frac{1}{\kappa_-^B} + \frac{1}{\kappa_+^B} - \frac{1}{\kappa_-^A} - \frac{1}{\kappa_+^A} \right ) \right ]^{-1}.
    \label{eq:Dk2nt}
\end{equation}
Here the smallest of the individual $\kappa_+^j$ or $\kappa_-^j$ tends to dominate and determine the sign of $K^\mathrm{ss}$.
The sign of $K^\mathrm{ss}$ is positive if the smallest coefficient is for the $B$ step, e.g. if the $B$ step has the higher ES barrier, so that $\kappa_-^B$ is smallest.
If there are no ES barriers, i.e. $\kappa_-^j = \kappa_+^j$, then the step with the smaller $\kappa_+^j$ determines the sign.
In this sub-case we have $R_0 = 0$, which simplifies Eq.~(\ref{eq:dfdta2}) for $df_\alpha/dt$.

For the sub-case of highly transparent steps, with $\kappa_0^j >> \kappa_+^j$ and $\kappa_-^j$, we have $\kapsq^j = \kappa_0^j (\kappa_+^j + \kappa_-^j)$ for both steps $j = A$ and $B$.
The expression for $K^\mathrm{ss}$ becomes a constant, independent of $f_\alpha$,
\begin{equation}
    K^\mathrm{ss} \approx
    \left ( \frac{1}{\kappa_-^B + \kappa_+^B} - \frac{1}{\kappa_-^A + \kappa_+^A} \right )^{-1}.
    \label{eq:Dk2t}
\end{equation}
Here the behavior just depends on the sums $\kappa_-^j + \kappa_+^j$ for each step.
It does not matter whether there are ES barriers;
the sign of $K^\mathrm{ss}$ is positive if $(\kappa_-^A + \kappa_+^A) > (\kappa_-^B + \kappa_+^B)$.

\subsection{Attachment-limited kinetics}

In the attachment-limited case, the final term is negligible in Eq.~(\ref{eq:Ra}) for $R_\alpha$ and in Eq.~(\ref{eq:Rb}) for $R_\beta$.
The coefficients $R_\alpha$ and $R_\beta$ become independent of $f_\alpha$.
The expression for $K^\mathrm{ss}$ is given by
\begin{equation}
    K^\mathrm{ss} \approx \big [ W_0^{al} + W_1^{al} (1 - 2f_\alpha) \big ]^{-1},
    \label{eq:Dk3}
\end{equation}
with coefficients
\begin{align}
    W_0^{al} & \equiv \frac{\kapsq^A - \kapsq^B + (\kappa_+^A + \kappa_-^A) \kappa_0^B - (\kappa_+^B + \kappa_-^B) \kappa_0^A}{(\kappa_+^B + \kappa_-^B) \kapsq^A + (\kappa_+^A + \kappa_-^A) \kapsq^B}, \label{eq:W0a} \\
    W_1^{al} & \equiv \frac{\kappa_+^B \kappa_+^A - \kappa_-^B \kappa_-^A}{(\kappa_+^B + \kappa_-^B) \kapsq^A + (\kappa_+^A + \kappa_-^A) \kapsq^B}. \label{eq:W1a}
\end{align}
The expression for $K^\mathrm{dyn}$ is independent of $f_\alpha$,
\begin{align}
    & K^\mathrm{dyn} \approx \left ( \left [ 
    \left ( \frac{\kappa_+^B}{\kapsq^B} + \frac{\kappa_-^A}{\kapsq^A} \right )^{-1} \right . \right . \nonumber \\
    &+ \left . \left . \left ( \frac{\kappa_-^B}{\kapsq^B} + \frac{\kappa_+^A}{\kapsq^A} \right )^{-1} \right ]^{-1}  
    +  \frac{\kappa_0^A}{\kapsq^A} + \frac{\kappa_0^B}{\kapsq^B} \right )^{-1} . \label{eq:dfdta3}
\end{align}
The diffusion coefficient $D$ does not enter into the solution for the attachment-limited case;
its role in the dynamics is taken by the combination of all the $\kappa$ coefficients given in Eq.~(\ref{eq:dfdta3}).
Since the denominators in Eqs.~(\ref{eq:W0a}-\ref{eq:W1a}) are always positive, the sign of $K^\mathrm{ss}$ is determined by the numerators.

For the sub-case of non-transparent steps, with $\kappa_0^A = \kappa_0^B = 0$, $\kapsq^j = \kappa_+^j \kappa_-^j$, the expressions for the coefficients in $K^\mathrm{ss}$ become
\begin{align}
    W_0^{al} & \equiv \frac{\kappa_+^A \kappa_-^A - \kappa_+^B \kappa_-^B}{(\kappa_+^B + \kappa_-^B) \kappa_+^A \kappa_-^A + (\kappa_+^A + \kappa_-^A) \kappa_+^B \kappa_-^B}, \label{eq:W0ant} \\
    W_1^{al} & \equiv \frac{\kappa_+^B \kappa_+^A - \kappa_-^B \kappa_-^A}{(\kappa_+^B + \kappa_-^B) \kappa_+^A \kappa_-^A + (\kappa_+^A + \kappa_-^A) \kappa_+^B \kappa_-^B}. \label{eq:W1ant}
\end{align}
This is the most complex sub-case.
Near $f_\alpha = 0.5$, the sign of $K^\mathrm{ss}$ is positive if $\kappa_+^B \kappa_-^B <  \kappa_+^A \kappa_-^A$.
At $f_\alpha > 0.5$, if the steps have normal ES barriers with $\kappa_-^j < \kappa_+^j$, the $W_1^{al}$ term will favor a negative sign.
Thus the sign of $K^\mathrm{ss}$ can change with $f_\alpha$.
The expression for $K^\mathrm{dyn}$ becomes
\begin{equation}
    K^\mathrm{dyn} \approx 
    \left ( \frac{1}{\kappa_-^B} + \frac{1}{\kappa_+^A} \right )^{-1} +  \left ( \frac{1}{\kappa_+^B} + \frac{1}{\kappa_-^A} \right )^{-1}. \label{eq:dfdta3nt}
\end{equation}
The dynamic coefficient has an interesting form, dominated by the terrace with the largest value of the \textit{smallest} attachment coefficient at its edges.

For the sub-case of highly transparent steps, with $\kappa_0^j >> \kappa_+^j$ and $\kappa_-^j$, $\kapsq^j = \kappa_0^j (\kappa_+^j + \kappa_-^j)$, the expression for $K^\mathrm{ss}$ becomes a constant identical to that for diffusion-limited kinetics with highly transparent steps,
\begin{equation}
    K^\mathrm{ss} \approx
    \left ( \frac{1}{\kappa_-^B + \kappa_+^B} - \frac{1}{\kappa_-^A + \kappa_+^A} \right )^{-1}.
    \label{eq:Dk3t}
\end{equation}
As before, the full-steady-state behavior just depends on the sums $\kappa_-^j + \kappa_+^j$ for each step.
The dynamics still differs from the diffusion-limited case, since the expression for $K^\mathrm{dyn}$ differs from Eq.~(\ref{eq:dfdta2}),
\begin{equation}
    K^\mathrm{dyn} \approx \left ( 
    \frac{1}{\kappa_-^B + \kappa_+^B} + \frac{1}{\kappa_-^A + \kappa_+^A} \right )^{-1} . \label{eq:dfdta3t}
\end{equation}

\subsection{Mixed kinetics}

The limits considered above assume that both terraces have the same type of kinetics, either diffusion- or attachment-limited, and that both steps have the same transparency, either zero or high. 
Because the attachment coefficients can be different for each step type, other limiting cases are possible.
Here we consider a particular mixed limit in which the $\kappa_+^A$ coefficient is much larger than the other five $\kappa_x^j$ (giving an $A$ step with a high ES barrier), and we asume that $\kappa_-^A + \kappa_0^A << D / w f_\alpha$.
We also assume that $\kappa_-^B << \kappa_+^B \kappa_0^B / (\kappa_+^B + \kappa_0^B)$, so that the $B$ step also has a high ES barrier.
In this case we have $\kapsq^A = \kappa_+^A (\kappa_-^A + \kappa_0^A)$ and $\kapsq^B = \kappa_+^B \kappa_0^B$.
The second and third terms in Eq.~(\ref{eq:Ra}) are negligible, giving $R_\alpha = (w/D)(\kappa_-^A + \kappa_0^A)$.
The second term in Eq.~(\ref{eq:Rb}) is negligible, giving $R_\beta = [ D/(w \kappa_0^B) + (1 - f_\alpha)]^{-1}$
and $R_\beta >> R_\alpha$.
The second terms in Eqs.~(\ref{eq:Sa}) and (\ref{eq:Sb}) are negligible, giving $S_\alpha = w^2 f_\alpha / (2D)$, $S_\beta = (w/2)(1-f_\alpha)/[D/w + (1-f_\alpha)\kappa_0^B]$.
The first terms in Eqs.~(\ref{eq:R0}) and (\ref{eq:S0}) are negligible, giving $R_0 = D/(w \kappa_+^B)$, $S_0 = -w/(2 \kappa_+^B)$.
This results in expressions
\begin{equation}
    K^\mathrm{ss} \approx \left [ 
    \frac{1}{\kappa_+^B} + \frac{(1 - 2 f_\alpha)}{\kappa_0^B} - \frac{w f_\alpha (1-f_\alpha)}{D} \right ]^{-1},
    \label{eq:Dkm}
\end{equation}
\begin{equation}
    K^\mathrm{dyn} \approx \left [ 
    \frac{1}{\kappa_+^B} + \frac{1}{\kappa_0^B} + \frac{w (1-f_\alpha)}{D} \right ]^{-1}.
    \label{eq:Dkmd}
\end{equation}
Even though $\kappa_+^A$ has the largest value, the sign of $K^\mathrm{ss}$ can be negative depending upon the relative size of the terms in Eq.~(\ref{eq:Dkm}).
It will be negative near $f_\alpha = 0.5$ for $D/\kappa_+^B < w/4$.
If $\kappa_0^B$ is small, it can become negative for $f_\alpha > 0.5$.

\begin{figure}
\includegraphics[width=0.85\linewidth]{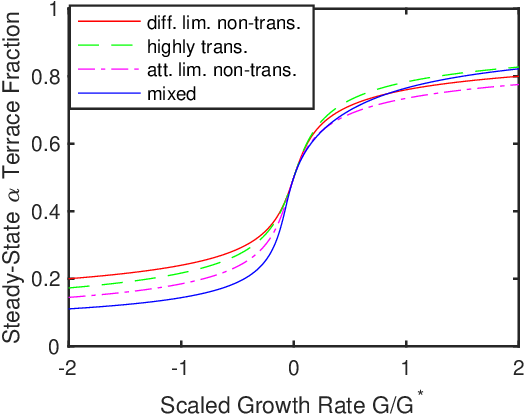}
\caption{Master curves of $f_\alpha^\mathrm{ss}$ vs. $G/G^*$ for different sub-cases: diffusion-limited kinetics with non-transparent steps, attachment-limited kinetics with non-transparent steps, either kinetics with highly transparent steps, and mixed kinetics. Parameter values used are given in Table \ref{tab:tableX}. \label{fig:master}}
\end{figure}

\begin{table}
\caption{ \label{tab:tableX} Parameter values used in BCF theory calculations for sub-cases shown in Fig.~\ref{fig:master}. All used $w = 5.73 \times 10^{-8}$ m, $\rho_0 = 1.13 \times 10^{19}$ m$^{-2}$, $\ell = 9.1 \times 10^{-10}$ m, $\rho_\mathrm{eq}^0 = 3.44 \times 10^{12}$ m$^{-2}$, $f_\alpha^0$ = 0.5, $M_0 = 0$.}
\begin{ruledtabular}
\begin{tabular}{ l || l | l | l | l | l}
Limited by: & diff. & diff. & attach. & attach. & mixed \\
\hline
Transparency: & zero & high & zero & high & mixed\\
\hline
$D$ ~~~ (m$^2$ s$^{-1}$) & $10^{-14}$ & $10^{-14}$ & $10^{-4}$ & $10^{-4}$ & $10^{-8}$ \\
$\kappa_+^A$ ~~ (m s$^{-1}$) & $10^2$ & $10^2$ & $10^2$ & $10^2$ & $10^6$ \\
$\kappa_-^A$ ~~ (m s$^{-1}$) & $10^1$ & $10^1$ & $10^1$ & $10^1$ & $10^{-3}$ \\
$\kappa_0^A$ ~~ (m s$^{-1}$) & $0$ & $10^3$ & $0$ & $10^3$  & $0$ \\
$\kappa_+^B$ ~~ (m s$^{-1}$) & $10^1$ & $10^1$ & $10^1$ & $10^1$ & $0.5$ \\
$\kappa_-^B$ ~~ (m s$^{-1}$) & $10^0$ & $10^0$ & $10^0$ & $10^0$ & $10^{-4}$ \\
$\kappa_0^B$ ~~ (m s$^{-1}$) & $0$ & $10^3$ & $0$ & $10^3$  & $1$ \\
\hline
$G^*$ ~~ (10$^{-3}$ ML/s) & $0.4$ & $1.2$ & $1.2$ & $1.2$ & $0.2$ \\
\end{tabular}
\end{ruledtabular}
\end{table}

\begin{table} 
\caption{ \label{tab:tableLP} Best-fit parameter values for the three limiting cases of the BCF model. \\ $\dagger$ Parameters obtained from diffusion-limited kinetics fits are inconsistent with the limiting approximation (see text). \\ $\dagger \dagger$ The A3 fit gives non-monotonic variation of $G^\mathrm{ss}(f_\alpha)$.}
\begin{tabular}{ c || c | c | c }
\hline
\hline
\multicolumn {4} {c} {Diffusion-limited kinetics} \\
\hline
Fit type: & D1 $\dagger$ & D2 $\dagger$ & D3 $\dagger$ \\
 & Fix $M_0 = 0$ & Fix $f_\alpha^0 = 0.5$ &  Vary Both \\
 & Vary $f_\alpha^0$ & Vary $M_0$ & $M_0$ and $f_\alpha^0$ \\
\hline
$R_0$ & $4.11 \times 10^{-2}$ & $6.61 \times 10^{-3}$ & $6.19 \times 10^{-1}$ \\
$D W_0^{dl}$ (m) & $2.13 \times 10^{-9}$ & $3.79 \times 10^{-10}$ & $1.71 \times 10^{-8}$ \\
$D W_1^{dl}$ (m) & $-3.89 \times 10^{-9}$ & $3.15 \times 10^{-9}$ & $-5.73 \times 10^{-8}$\\
$D \rho_\mathrm{eq}^0 \ell^3$ (m$^3$ s$^{-1}$)& $7.85 \times 10^{-24}$ & $4.09 \times 10^{-24}$ & $4.32 \times 10^{-23}$   \\
$f_\alpha^0$ & $0.388$ & $0.500$ & $0.286$ \\
$D \rho_\mathrm{eq}^0 M_0$ (s$^{-1}$)& $0$ & $3.93 \times 10^{-2}$ & $-4.52 \times 10^{-1}$   \\
\hline
$f_\alpha^*$ & $0.388$ & $0.377$ & $0.453$ \\
$G^*$ (ML/s) & $0.0027$ & $0.0025$ & $0.0067$ \\
$t^*$ (s) & $4900$ & $4910$ & $2530$ \\
$(w/\ell)^3 M_0$ & $0$ & $1.81$ & $-1.97$ \\
\hline
$\chi^2$ & $60.4$ & $101.6$ & $25.3$ \\
\hline
\hline
\multicolumn {4} {c} {Attachment-limited kinetics} \\
\hline
Fit type: & A1 & A2 &  A3 $\dagger\dagger$ \\
 & Fix $M_0 = 0$ & Fix $f_\alpha^0 = 0.5$ &  Vary Both \\
 & Vary $f_\alpha^0$ & Vary $M_0$ & $M_0$ and $f_\alpha^0$ \\
\hline
$K^\mathrm{dyn} W_0^{al}$ & $8.29 \times 10^{-2}$ & $7.88 \times 10^{-2}$ & $7.94 \times 10^{-2}$ \\
$K^\mathrm{dyn} W_1^{al}$ & $3.27 \times 10^{-2}$ & $7.59 \times 10^{-2}$ & $9.60 \times 10^{-2}$ \\
$K^\mathrm{dyn} \rho_\mathrm{eq}^0 \ell^3$ & $5.06 \times 10^{-16}$ & $2.85 \times 10^{-16}$ & $2.04 \times 10^{-16}$  \\
(m$^2$ s$^{-1}$) & & & \\
$f_\alpha^0$ & $0.379$ & $0.500$ & $0.584$  \\
$K^\mathrm{dyn} \rho_\mathrm{eq}^0 M_0$ & $0$ & $2.52 \times 10^6$ & $3.14 \times 10^6$  \\
(m$^{-1}$ s$^{-1}$) & & &  \\
\hline
$f_\alpha^*$ & $0.379$ & $0.384$ & $0.396$ \\
$G^*$ (ML/s) & $0.0023$ & $0.0019$ & $0.0018$ \\
$t^*$ (s) & $4720$ & $5400$ & $5720$ \\
$(w/\ell)^3 M_0$ & $0$ & $1.66$ & $2.90$ \\
\hline
$\chi^2$ & $72.9$ & $51.9$ & $50.2$ \\
\hline
\hline
\multicolumn {4} {c} {Mixed kinetics} \\
\hline
Fit type: & M1 & M2 & M3 \\
 & Fix $M_0 = 0$ & Fix $f_\alpha^0 = 0.5$ &  Vary Both \\
 & Vary $f_\alpha^0$ & Vary $M_0$ & $M_0$ and $f_\alpha^0$ \\
\hline
$D / \kappa_+^B$ (m)& $1.82 \times 10^{-8}$ & $1.90 \times 10^{-8}$ & $1.80 \times 10^{-8}$  \\
$D / \kappa_0^B$ (m)& $9.09 \times 10^{-9}$ & $1.19 \times 10^{-8}$ & $7.67 \times 10^{-9}$ \\
$D \rho_\mathrm{eq}^0 \ell^3$ (m$^3$ s$^{-1}$)& $3.58 \times 10^{-23}$ & $3.25 \times 10^{-23}$ & $3.87 \times 10^{-23}$  \\
$f_\alpha^0$ & $0.443$ & $0.500$ & $0.416$ \\
$D \rho_\mathrm{eq}^0 M_0$ (s$^{-1}$)& $0$ & $8.16 \times 10^{-2}$ & $-6.40 \times 10^{-2}$   \\
\hline
$f_\alpha^*$ & $0.443$ & $0.461$ & $0.442$ \\
$G^*$ (ML/s) & $0.0028$ & $0.0024$ & $0.0032$ \\
$t^*$ (s) & $4150$ & $4540$ & $3840$ \\
$(w/\ell)^3 M_0$ & $0$ & $0.47$ & $-0.31$ \\
\hline
$\chi^2$ & $25.0$ & $30.3$ & $24.1$ \\
\hline
\hline
\end{tabular}
\end{table}

\subsection{Summary of limiting cases}

While there are 9 free parameters in the full simplified analytical solution, in the limiting cases considered above the number of effective parameters is smaller, since only certain combinations of $D / \kappa_x^j$ enter the solutions.
The diffusion-limited kinetics solutions reduce these 6 to 3 combinations, leaving a total of 6 unknown quantities.
The sub-cases of non-transparent or highly transparent steps reduce the number of effective parameters by one or two more.
The attachment-limited kinetics solutions reduce these 6 to 2 combinations, leaving a total of 5 unknown quantities.
The highly transparent sub-case reduces this by one.
The mixed kinetics solution reduces these 6 to 2 combinations, leaving a total of 5 unknown quantities, $D/\kappa_+^B$, $D/\kappa_0^B$, $D \rho_\mathrm{eq}^0 M_0$, $D \rho_\mathrm{eq}^0 \ell^3$, and $f_\alpha^0$.

Figure~\ref{fig:master} shows some examples of master curves of $f_\alpha^\mathrm{ss}$ vs. $G/G^*$, calculated using the simplified analytical solution Eqs.~(\ref{eq:Dk}-\ref{eq:Gstar}) with parameter values given in Table \ref{tab:tableX}.
These correspond to the limiting cases discussed above.
Curves for both the diffusion-limited cases (non- or highly transparent) have inversion symmetry around $G = 0$, $f_\alpha^\mathrm{ss} = 0.5$, reflecting the symmetry of Eqs.~(\ref{eq:Mfa}), (\ref{eq:fSSG}), and (\ref{eq:Dk2}) when $M_0 = 0$ and $f_\alpha^0 = 0.5$.
The curves for both highly transparent cases (diffusion- or attachment-limited) are identical.
The attachment-limited non-transparent case is not symmetric, reflecting the $W_1^{al}$ term in Eq.~(\ref{eq:Dk3}).
The mixed case is least symmetric, and its shape depends in a complex way on the relative sizes of the terms in Eq.~(\ref{eq:Dkm}).
When $M_0$ and $f_\alpha^0$ values are used that deviate from $0$ and $0.5$, respectively, all curves become asymmetric.

\section{Limiting-Case Fits}

To understand how well the measurements constrain the BCF model parameters and the physics underlying them, we searched for the best fit using the expressions obtained in Appendix B for each of the three limiting cases (diffusion-limited, attachment-limited, and mixed kinetics).
Table~\ref{tab:tableLP} summarizes the best-fit values of the parameters obtained, and also gives the characteristic values $f_\alpha^*$, $G^*$, $t^*$, and $(w/\ell)^3 M_0$ for each fit.
Nine fits were carried out.
For each of the three limiting cases, we considered three functional forms for $M$: fixed $M_0 = 0$ with varying $f_\alpha^0$; fixed $f_\alpha^0 = 0.5$ with varying $M_0$; and varying both $M_0$ and $f_\alpha^0$.
The fits are labelled with a letter (D, A, or M) indicating the kinetic limit and a number (1, 2, or 3) corresponding to the form used for $M$.

The fits using the mixed kinetics limit (M1, M2, and M3) generally give better results (lower $\chi^2$) than the diffusion- or attachment-limited cases.
The best fit obtained previously \cite{2020_Ju_NC} using a single relaxation time for each transition is similar to the M1 fit. 
However, here the diffusion-limited case with variable $M_0$ and $f_\alpha^0$ (D3) gives a similar low value of $\chi^2$.
All of the fits except one (A3) produce $G^\mathrm{ss}(f_\alpha)$ that increase monotonically.
A non-monotonic $G^\mathrm{ss}(f_\alpha)$ is problematic because it leads to non-unique $f_\alpha^\mathrm{ss}(G)$ values needed for comparison with the experiments.
(For A3 we used only the monotonic portion of the $G^\mathrm{ss}(f_\alpha)$ curve up to the first maximum to obtain $f_\alpha^\mathrm{ss}(G)$.)
Another issue arises with the diffusion-limited kinetics fits.
The parameter sets from the fits are not self-consistent with the approximations used to obtain the expressions for this limit, which require $|D W_1^{dl}| << w = 5.73 \times 10^{-8}$~m.
The D3 fit most clearly violates this approximation.


\bibliography{bibliography/BCF_ABsteps_shortJnames}

\end{document}


\title{Supplemental Material for ``Burton-Cabrera-Frank theory for surfaces with alternating step types''}

\author{Guangxu Ju}
    \email[correspondence to: ]{juguangxu@gmail.com}
	\altaffiliation[current address: ]{Lumileds Lighting Co., San Jose, CA 95131 USA.}
	\affiliation{Materials Science Division, Argonne National Laboratory, Lemont, IL 60439 USA}
\author{Dongwei Xu}
	\affiliation{Materials Science Division, Argonne National Laboratory, Lemont, IL 60439 USA}
	\affiliation{School of Energy and Power Engineering, Huazhong University of Science and Technology, Wuhan, Hubei 430074, China}
\author{Carol Thompson}
	\affiliation{Department of Physics, Northern Illinois University, DeKalb, IL 60115 USA}
\author{Matthew J. Highland}
	\affiliation{X-ray Science Division, Argonne National Laboratory, Lemont, IL 60439 USA}
\author{Jeffrey A. Eastman}
	\affiliation{Materials Science Division, Argonne National Laboratory, Lemont, IL 60439 USA}
\author{Weronika Walkosz}
	\affiliation{Department of Physics, Lake Forest College, Lake Forest, IL 60045 USA}
\author{Peter Zapol}
	\affiliation{Materials Science Division, Argonne National Laboratory, Lemont, IL 60439 USA}
\author{G. Brian Stephenson}
    \email[correspondence to: ]{stephenson@anl.gov}
	\affiliation{Materials Science Division, Argonne National Laboratory, Lemont, IL 60439 USA}

\date{January 26, 2022}  

\begin{abstract}
This Supplemental Material contains checks of self-consistency of assumptions in the derivations and a description of the extraction of the three characteristic times from experimental measurements of transitions.
\end{abstract}

\maketitle

\section{Self-consistency checks}

\subsection{Neglect of advective terms}

We wish to compare the contributions to the adatom flux $J_i$ of the diffusive and advective terms $-D \nabla_y \rho_i$ and $-v \rho_i$, respectively, where $v = (v_A + v_B)/2$ is the average step velocity.
First consider the simplified analytical solution of Section II.E in the main text, which neglects the advective terms.
The diffusive flux divergence term in the continuity equation is a constant, $-\nabla_y \cdot J_i = D \nabla_y^2 \rho_i = -\rho_0 G$ for both terraces $i = \alpha$ or $\beta$.
The flux $J_i$ changes linearly from $J_i^-$ to $J_i^+$ across each terrace.
The difference across a terrace is
\begin{equation}
    J_i^+ - J_i^- = w_i \nabla_y \cdot J_i = w_i \rho_0 G.
\end{equation}
The sum of the flux differences across alpha and beta terraces is
\begin{equation}
    J_\alpha^+ - J_\alpha^- + J_\beta^+ - J_\beta^- = \rho_0 G w. \label{eq:adv}
\end{equation}
To compare this to the advective term, we use Eq.~(28) of the main text, relating the sum of the step velocities to the net growth rate and step spacing by $v_A + v_B = G w$, giving an advective term $-v \rho_i = -\rho_i G w / 2$.
Comparing this to Eq.~(\ref{eq:adv}), one can see that adding the advective term to all fluxes has a negligible effect when the adatom coverage is small, $\rho_i << \rho_0$.

\subsection{Quasi-steady-state approximation}

Using the simplified analytical solution, we can develop criteria for the self-consistency of the quasi-steady-state approximation used to solve Eq.~(1) in the main text.
In this case the evaporation and deposition terms on the right-hand side of Eq.~(1) simply sum to $\rho_0 G$, so the quasi-steady-state approximation can be expressed as
\begin{equation}
    \left | \frac{\partial \rho_i}{\partial t} \right | << | \rho_0 G |. \label{eq:qsscrit}
\end{equation}
We can solve for the left-hand side using the quasi-steady-state solution, Eq.~(8) in the main txt, where the time derivative is due solely to the evolution of $f_\alpha$,
\begin{equation}
    \frac{\partial \rho_i}{\partial t} = \left . \frac{\partial \rho_i}{\partial f_\alpha} \right |_y \, \, \frac{d f_\alpha}{dt}.
\end{equation}
For the simplified analytical solution, the first two terms in Eq.~(8) do not depend on $f_\alpha$, giving
\begin{equation}
    \left . \frac{\partial \rho_i}{\partial f_\alpha} \right |_y = \frac{\partial C_{2i}}{\partial f_\alpha} \, \, \frac{y}{\sqrt{D \tau}}.
\end{equation}
The maximum amplitude occurs at the terrace boundaries, $y = \pm w_i / 2$.
At the positive boundary we have
\begin{equation}
    \frac{\partial \rho_i}{\partial t} = \frac{f_i \, K^\mathrm{dyn} (G - G^\mathrm{ss})}{2 K^\mathrm{ss}} \, \, \frac{\partial [I_i \, R_i \, (\rhot_\mathrm{eq}^A - \rhot_\mathrm{eq}^B) + S_i \, \rho_0 G ]}{\partial f_\alpha},  \label{eq:drhoidt}
\end{equation}
where we have used Eqs.~(51-52) and Eq.~(68) of the main text, and defined the sign factor $I_i = +1$ for $i = \alpha$ and $I_i = -1$ for $i = \beta$.
To evaluate the derivative on the right, we can start by differentiating Eqs.~(53-56) of the main text to give
\begin{align}
    \frac{d R_i}{d f_\alpha} &= -I_i \, R_i^2, \\
    \frac{d S_i}{d f_\alpha} &= -I_i \, S_i \left( R_i - \frac{1}{f_i} \right ).
\end{align}

Using Eq.~(59) and some algebra, we can evaluate the derivative on the right of Eq.~(\ref{eq:drhoidt}).
It has a linear dependence on $\rho_0 G$,
\begin{equation}
    \frac{\partial [ I_i \, R_i \, (\rhot_\mathrm{eq}^A - \rhot_\mathrm{eq}^B) + S_i \, \rho_0 G ]}{\partial f_\alpha} = H_0^i + H_1^i \, \rho_0 G,
\end{equation}
where the coefficients are
\begin{align}
    &H_0^i = 2 \rho_\mathrm{eq}^0 I_i \, R_i \\
    &\times \left ( \frac{M^\prime - I_i \, R_i M}{1 + R_0 (R_\alpha + R_\beta)} 
    + \frac{R_0 M (R_\alpha^2 - R_\beta^2)}{[1 + R_0 (R_\alpha + R_\beta)]^2} \right ), \nonumber \\
    &H_1^i = -I_i \, S_i \left (R_i - \frac{1}{f_i} \right ) - \frac{R_i^2 [S_0 + R_0 (S_\beta - S_\alpha)]}{1 + R_0 (R_\alpha + R_\beta)}  \\
    &+ R_0 \, I_i \, R_i \left ( \frac{ R_0 [S_\beta - S_\alpha - S_0 (R_\alpha +R_\beta)] (R_\alpha^2 - R_\beta^2)}{[1 + R_0 (R_\alpha + R_\beta)]^2} \right .  \nonumber \\
    &\left . + \frac{ S_\alpha \left( R_\alpha - \frac{1}{f_\alpha} \right ) + S_\beta \left( R_\beta - \frac{1}{f_\beta} \right ) + S_0 (R_\alpha^2 - R_\beta^2)}{1 + R_0 (R_\alpha + R_\beta)} \right ). \nonumber
\end{align}

Substituting these into Eq.~(\ref{eq:drhoidt}) gives a quadratic dependence on $\rho_0 G$,
\begin{equation}
    \frac{\partial \rho_i}{\partial t} = h_0^i + h_1^i \rho_0 G + h_2^i (\rho_0 G)^2, \label{eq:drhodt2}
\end{equation}
where the coefficients are
\begin{align}
    h_0^i &= -H_0^i \, \frac{2 f_i \, M \, \rho_\mathrm{eq}^0 D (R_\alpha + R_\beta)}{w^2 \rho_0 (1 + R_0 (R_\alpha + R_\beta)},\\
    h_1^i &= H_0^i \, \frac{f_i \, D \, [S_\beta - S_\alpha - S_0(R_\alpha + R_\beta)]}{\rho_0 w^2 (1 + R_0 (R_\alpha + R_\beta)}, \nonumber \\
    &- H_1^i \, \frac{2 f_i \, M \, \rho_\mathrm{eq}^0 \, D (R_\alpha + R_\beta)}{w^2 \rho_0 (1 + R_0 (R_\alpha + R_\beta)},\\
    h_2^i &= H_1^i \, \frac{f_i \, D \, [S_\beta - S_\alpha - S_0(R_\alpha + R_\beta)]}{\rho_0 w^2 (1 + R_0 (R_\alpha + R_\beta)}.
\end{align}

We can separately consider the effect of the constant, linear, and quadratic terms in Eq.~(\ref{eq:drhodt2}) on the quasi-steady-state criterion Eq.~(\ref{eq:qsscrit}).
The constant term imposes a lower limit on the growth rate that satisfies the criterion, $|h_0^i / \rho_0| << |G|$, effectively giving an accuracy value for $G^\mathrm{ss}(f_\alpha)$.
The linear term must obey $|h_1^i| << 1$ to satisfy the criterion.
The quadratic term imposes an upper limit on the growth rate that satisfies the criterion, $|G| << |1/\rho_0 h_2^i|$.
All of the terms are functions of $f_\alpha$.

We have calculated these values for all the parameter sets given in the fits in Table~III in the main text.
In all cases the accuracy of $G^\mathrm{ss}(f_\alpha)$ given by $|h_0^i / \rho_0|$ is better than $10^{-5}$~ML/s for terrace fractions between $0.02 < f_\alpha < 0.98$.
Likewise in all cases $|h_1^i| << 10^{-7}$ over this range, and the maximum $|G|$ given by $|1/\rho_0 h_2^i|$ is greater than $10^{7}$~ML/s for all values of $f_\alpha$.

\subsection{Simplified analytical solution}

Here we check the self-consistency of the assumptions made in deriving the simplified analytical solution.
One assumption is that the adatom lifetime $\tau$ is sufficiently large that the adatom diffusion length $\sqrt{D \tau}$ is much larger than both the terrace width $w$ (so that $c_i \approx 1$ and $s_i \approx w_i / 2 \sqrt{D \tau}$) and the appropriate kinetic lengths (so that $\pt_A \qt_B >> 1$ and $\pt_B \qt_A >> 1$).
A second assumption is that the $\rhot_\mathrm{eq}^j$ do not differ much from $\rho_\mathrm{eq}^0$.
This requires that $|M| << 1$ (so that $\rho_\mathrm{eq}^j \approx \rho_\mathrm{eq}^0$), and that the second term in Eq.~(42) in the main text is negligible with respect to $\rho_\mathrm{eq}^0$, or
\begin{equation}
    \left | \frac{v_j \rho_0 \kappa_0^j}{\kapsq^j} \right | << \rho_\mathrm{eq}^0 \label{eq:vjlim}
\end{equation}
for both steps $j = A$ and $B$.
To evaluate this we can write the expressions for the step velocities Eqs.~(57-58) as
\begin{align}
    v_A &= \frac{w G}{2} + \frac{2 \, M \, K^\mathrm{dyn} \rho_\mathrm{eq}^0}{\rho_0} \left [ \frac{G - G^\mathrm{ss}}{G^\mathrm{ss}} \right ],  \label{eq:vas2}\\
    v_B &= \frac{w G}{2} - \frac{2 \, M \, K^\mathrm{dyn} \rho_\mathrm{eq}^0}{\rho_0} \left [ \frac{G - G^\mathrm{ss}}{G^\mathrm{ss}} \right ]. \label{eq:vbs2}
\end{align}
The first term gives the full-steady-state velocity, and the second term gives the difference in velocity when $f_\alpha$ differs from $f_\alpha^\mathrm{ss}$.
For the full-steady-state term, relation (\ref{eq:vjlim}) gives maximum growth rate magnitudes of
\begin{equation}
    |G| << G_{max}^j \equiv \frac{2 \rho_\mathrm{eq}^0 \kapsq^j}{w \rho_0 \kappa_0^j}  \label{eq:Gmax}
\end{equation}
for both steps $j = A$ and $B$.
For the dynamic term, relation (\ref{eq:vjlim}) gives maximum growth rate difference magnitudes of
\begin{equation}
    |G - G^\mathrm{ss}| << \Delta G_{max}^j \equiv \left | \frac{\rho_0 G^\mathrm{ss} \kapsq^j}{2 K^\mathrm{dyn} M \rho_\mathrm{eq}^0 \kappa_0^j} \right | . \label{eq:DGmax}
\end{equation}

\begin{figure}
\includegraphics[width=0.8\linewidth]{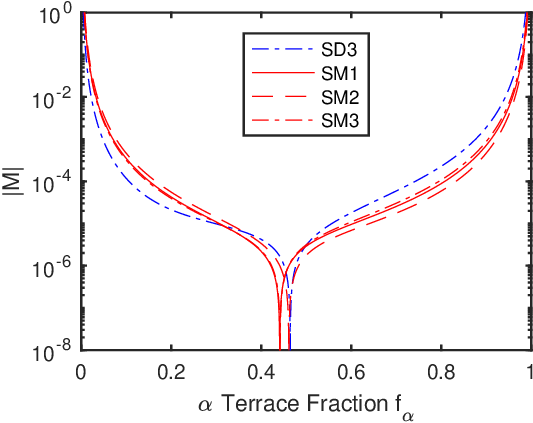}
\caption{Curves show step chemical potential magnitude $|M|$ as a function of $f_\alpha$ obtained from BCF simplified analytical fits shown in Table III. This quantity must be much less than unity to satisfy the simplified analytical approximation.\label{fig:M}}
\end{figure}

Using the parameter sets obtained from the simplified analytical fits, we have tested all these criteria for the validity of the simplified analytical solution.
In addition to the fit parameter values, we need values of the equilibrium adatom density $\rho_\mathrm{eq}^0$, the adatom diffusivity $D$, and the adatom lifetime $\tau$.
We use estimates extracted from the literature of $\rho_\mathrm{eq}^0 = 3.44 \times 10^{12}$~m$^{-2}$, $D = 1.35 \times 10^{-8}$ m$^2$~s$^{-1}$, and $\tau = 1.7 \times 10^{-4}$~s \cite{2020_Ju_NC}.
These give an adatom diffusion length of $\sqrt{D \tau} = 1.5 \times 10^{-6}$~m.
This is much larger than the terrace width $w = 5.73 \times 10^{-8}$~m, and also large enough to satisfy $\pt_A \qt_B >> 1$ and $\pt_B \qt_A >> 1$ for all cases.
Figure~\ref{fig:M} shows $M$ as a function of $f_\alpha$ for the four best fits.
The criterion $M << 1$ is violated only for ranges of $f_\alpha$ near zero and unity.
To test the criteria $|G| << G_{max}^j$ from equation Eq.~(\ref{eq:Gmax}), we calculated values for $G_{max}^j$ from the fit parameters using the estimated values of $\rho_\mathrm{eq}^0$ and $D$.
The maximum growth rate of $0.0127$~ML/s satisfies the criteria for all fits.
Likewise, the criteria $|G - G^\mathrm{ss}| << \Delta G_{max}^j$ from Eq.~(\ref{eq:DGmax}) are satisfied for all cases at all $f_\alpha$.

\section{Extraction of characteristic times from transition measurements}

\begin{figure}
\includegraphics[width=0.9\linewidth]{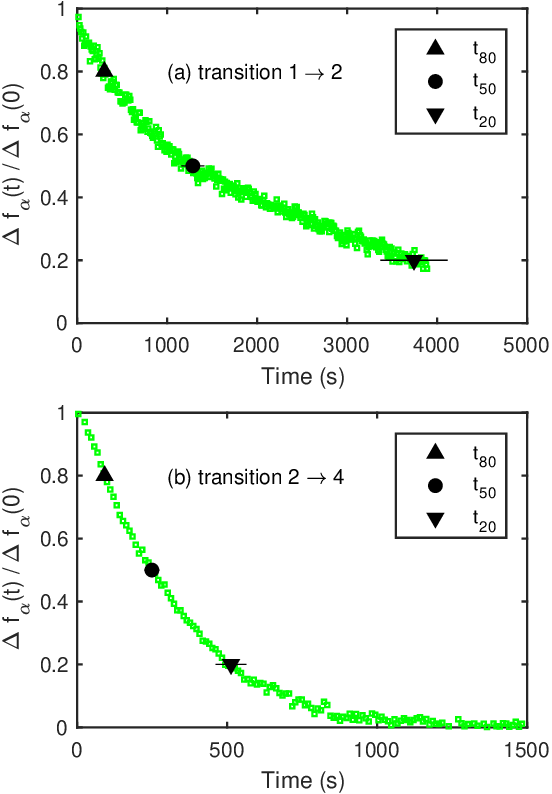}
\caption{Green points show normalized deviation of $\alpha$ terrace fraction from its steady-state value as a function of time, from measurement of transition from growth condition (a) 1 to 2, (b) 2 to 4. Large symbols show extracted values of $t_{80}$, $t_{50}$, and $t_{20}$ given in Table~II of the main text.\label{fig:tx12}}
\end{figure}

To account for potentially non-exponential relaxation upon changing conditions, we extracted three characteristic times from the observed $f_\alpha(t)$, rather than the single relaxation time used previously \cite{2020_Ju_NC}.
The values are given in Table~\ref{tab:tableZ} and Table~II of the main text.
The three characteristic times, $t_{80}$, $t_{50}$, and $t_{20}$, are the times for the normalized deviation of the terrace fraction from its steady-state value, $\Delta f_\alpha(t) / \Delta f_\alpha(0)$, to reach 80\%, 50\%, and 20\%, respectively, after a change of growth rate at $t = 0$.
Figure \ref{fig:tx12} illustrates how these values were obtained from the measured data.

\begin{acknowledgments}
  Work supported by the U.S Department of Energy (DOE), Office of Science, Office of Basic Energy Sciences, Materials Science and Engineering Division.
  Measured values shown are from experiments at beamline 12ID-D of the Advanced Photon Source, a DOE Office of Science user facility operated by Argonne National Laboratory.
\end{acknowledgments}

\begin{table} 
\caption{ \label{tab:tableZ} Measured values for OMVPE growth of GaN $(0 0 0 1)$ at 1073~K \cite{2020_Ju_NC}. Top: steady-state terrace fraction $f_\alpha$ under four growth conditions defined by the deposition flux $F$, evaporation flux $\rho_\mathrm{eq}^0 / \tau$, and the resulting net growth rate $G$. Bottom: characteristic times for relaxation of $f_\alpha(t)$ upon changing conditions.}
\begin{tabular}{ c || c | c | c | c }
\hline
\hline
\multicolumn {5} {c} {Steady-State} \\
\hline
Condition & 1 & 2 & 3 & 4 \\
\hline
$F$ (m$^{-2}$ s$^{-1}$) & $0$ & $0$ & $1.43 \times 10^{17}$ & $1.43 \times 10^{17}$ \\
$\rho_\mathrm{eq}^0 / \tau$ (m$^{-2}$ s$^{-1}$) & $2.0 \times 10^{16}$ & $0$ & $2.0 \times 10^{16}$ & $0$ \\
$G$ (ML/s) & $-0.0018$ & $0.0000$ & $0.0109$ & $0.0127$ \\
\hline
$f_\alpha^\mathrm{ss}$ & $0.111$ & $0.461$ & $0.811$ & $0.868$ \\
& $\pm 0.013$ & $\pm 0.018$ & $\pm 0.014$ & $\pm 0.011$ \\
\hline
\hline
\multicolumn {5} {c} {Transition Dynamics} \\
\hline
Transition & \multicolumn {2} {c |} {$1 \rightarrow 2$} & \multicolumn {2} {c} {$2 \rightarrow 4$} \\
\hline
$t_{80}$ (s) & \multicolumn {2} {c |} {$300 \pm 30$} & \multicolumn {2} {c} {$92 \pm 9$} \\
$t_{50}$ (s) & \multicolumn {2} {c |} {$1290 \pm 130$} & \multicolumn {2} {c} {$250 \pm 25$} \\
$t_{20}$ (s) & \multicolumn {2} {c |} {$3740 \pm 370$} & \multicolumn {2} {c} {$510 \pm 50$} \\
\hline
\hline
\end{tabular}
\end{table}

\clearpage


\bibliography{bibliography/BCF_ABsteps_shortJnames}